\documentclass[12pt]{article}
\usepackage[numbers]{natbib}
\usepackage{graphicx}

\newcommand{\rd}{{\rm d}}
\newcommand{\bm}[1]{\mbox{\protect\boldmath$#1$}}

\begin{document}

\title{Magnetic neutral points and electric lines of force in strong gravity of a rotating 
black hole\footnote{To appear in International Journal of Astronomy and Astrophysics}}

\author{V. Karas, O. Kop\'a\v{c}ek, and D. Kunneriath\\[10pt]
Astronomical Institute, Academy of Sciences,\\ Bo\v{c}n\'{\i}~II~1401, CZ-14100~Prague, Czech Republic\\
E-mail: vladimir.karas@cuni.cz}

\maketitle

\begin{abstract}
Magnetic field can be amplified and twisted near a
supermassive black hole residing in a galactic nucleus.
At the same time magnetic null points develop near the 
horizon. We examine a large-scale
oblique magnetic field near a rotating (Kerr) black hole as an origin of 
magnetic layers, where the field direction changes abruptly in
the ergosphere region.
In consequence of this, magnetic null points can develop by purely
geometrical effects of the strong gravitational field and the
frame-dragging mechanism. We identify magnetic nulls as possible sites of 
magnetic reconnection and suggest that particles may be accelerated
efficiently by the electric component. The situation we discuss 
is relevant for starving nuclei of some
galaxies which exhibit episodic accretion events, namely,
Sagittarius~A* black hole in our Galaxy.
\end{abstract}

%\keywords{galaxies: nuclei --- Galaxy: center --- black hole physics --- magnetic fields}

\section{Introduction}
Most galaxies including the Milky Way are believed to host a
supermassive black hole in the centre \citep{ess05,m07}. The black hole is embedded in 
a surrounding gaseous medium and magnetic fields and it can often rotate
with an almost extreme value of angular momentum, thereby interacting
with magnetic fields of external origin \citep{bw13}.
Electromagnetic processes have been proposed to explain 
the origin of high-energy particles and radiation flares from black hole
(BH) sources. However, there is a variety of possible mechanisms whose
details have not yet been fully understood \citep{bbr84,ess05,ge10,m07}. 
Here we consider the innermost regions where a
cooperation of strong gravitational and electromagnetic fields seems to
be required \citep{bg99,l00}. We study narrow magnetic layers arising
by a purely gravitational effect of a rotating BH. These layers are 
potential sites of acceleration which can operate as the
electric field acts across magnetic null points.

We begin this investigation by assuming an organised (ordered) 
large-scale magnetic field. This
is obviously a crude starting point, but a sensible one, 
representing the field generated by sources distant from the BH 
(astrophysical black holes are practically uncharged and possess 
no intrinsic dipole-like magnetic fields). Such a premise about the 
field structure appears adequate also in the case of Sagittarius~A* 
(Sgr~A*), where the black hole of $4\times 10^6$ solar masses resides 
\citep{ge10,m06}. Given the compact size of the black hole horizon,
the magnetic field generated by external sources appears to be
effectively uniform on the length-scale a several gravitational
radii. However, the field intensity is uncertain. On large 
scales (i.e., greatly exceeding the gravitational radius) the field should not go
beyond a few milligauss  \citep{mud06,mbb08}, while on medium scales
($\simeq10$--$20\,r_{\rm{}g}$) it might be amplified to tens of gauss
\citep{dl94,ebz08,mfy01}. 

Magnetic reconnection occurs when the magnetic field lines change their
connectivity \citep{pf00,som06}. This happens as
topologically distinct regions approach each other. The standard setup
involves the violation of the ideal MHD
approximation just on the boundary between neighbouring magnetic domains
where the field direction changes rapidly. 

One can ask if the BH proximity creates conditions favourable to 
incite reconnection, leading to plasma heating and particle acceleration. This
could generate the flaring activity \citep{bbb01,gso03,ppa03}. The typical rise 
time of Sgr~A* flares lasts several minutes, i.e.\ a fraction of the orbital 
period near the innermost stable circular orbit (ISCO). Variety of processes 
have been considered for Sgr~A* radiation. For example, \citet{lmp06a,lmp06b} 
propose that stochastic acceleration of electrons in the 
turbulent magnetic field is responsible for the
submillimeter emission within $\simeq20r_{\rm{}g}$. Also,
\citet{bml07} elaborate on the idea that Sgr~A* may be an
important site for particle acceleration. We will discuss a
complementary scheme of a magnetically dominated system. 

We show that antiparallel field lines are brought into mutual contact,
within the low-density conditions, by the frame-dragging
(gravito-magnetic) action of the rotating BH. One expects that a
dissipation region develops where the magnetic field structure changes
abruptly across a separatrix curve, so these spots, occurring just above
the ISCO, can act as places where particles are energised
\citep{dl94}. Ingredients necessary for this scenario to work -- i.e.\
an ordered magnetic field due to external sources plus the diluted
plasma environment of disturbed stars -- are naturally present near
Sgr~A* black hole. In addition to our previous work \citep{k08} we
show also the electric field threading the magnetic nulls. Thereby 
the electric component is capable of accelerating the charged matter
once it is injected in the area of the magnetic null.

\section{Forming magnetic layers by gravitational frame-dragging}
In underluminous galactic nuclei, it is likely that plasma is only
episodically injected into the central region, perhaps by passing stars
gradually sinking down to the BH. We model the gravitational
field by Kerr metric
\citep{mtw73}; we will use geometrical units with $c=G=1$ and
scale all quantities by the BH mass, $M$. The
gravitational radius is 
$r_{\rm{}g}=c^{-2}GM\approx4.8\times10^{-7}M_7\,$pc, and the
corresponding light-crossing time-scale
$t_{\rm{}g}=c^{-3}GM\approx49\,M_7\,$sec, where $M_7\equiv
M/(10^7M_\odot)$. 

\begin{figure*}[tbh]
\includegraphics[width=.49\textwidth]{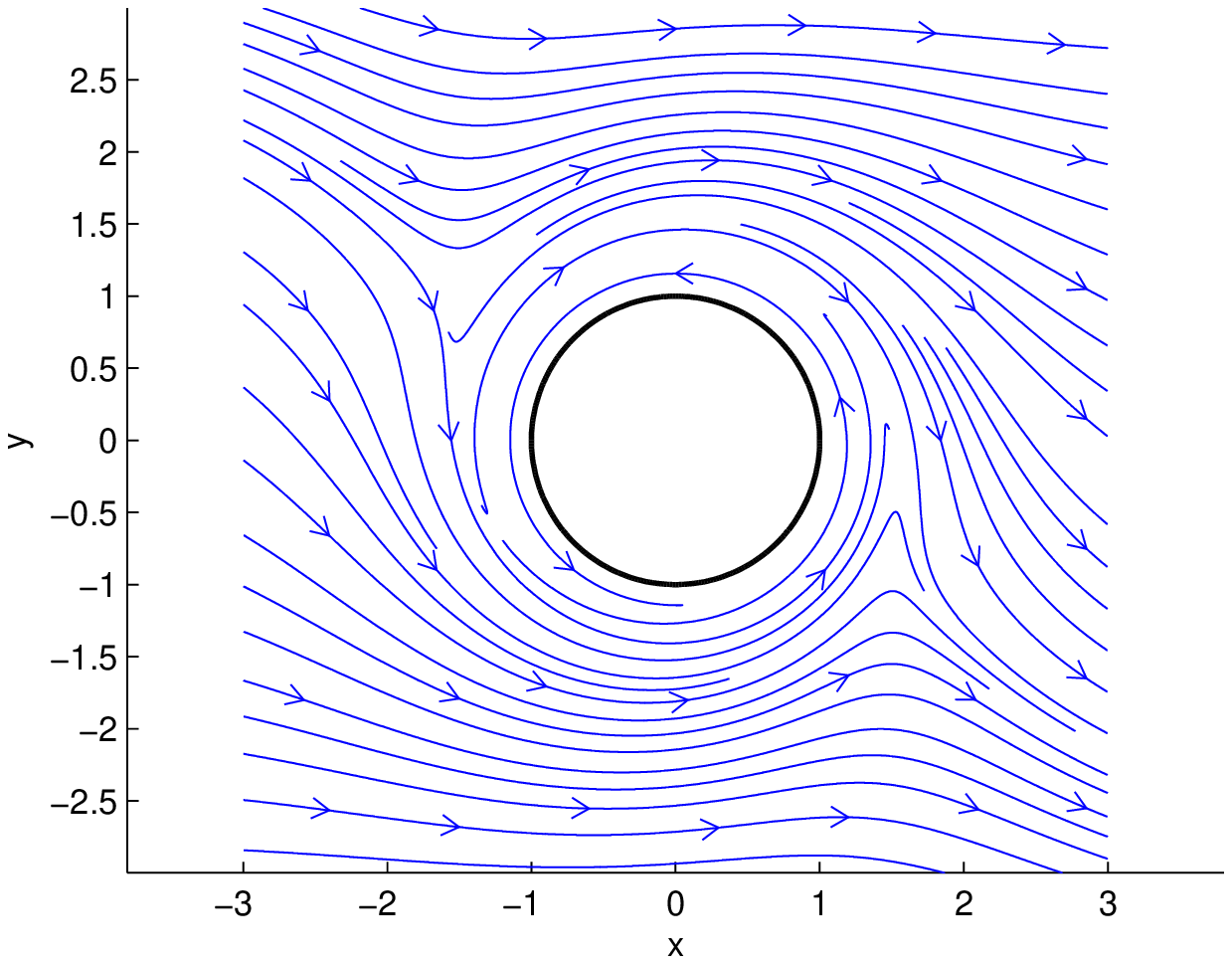}
\hfill
\includegraphics[width=.49\textwidth]{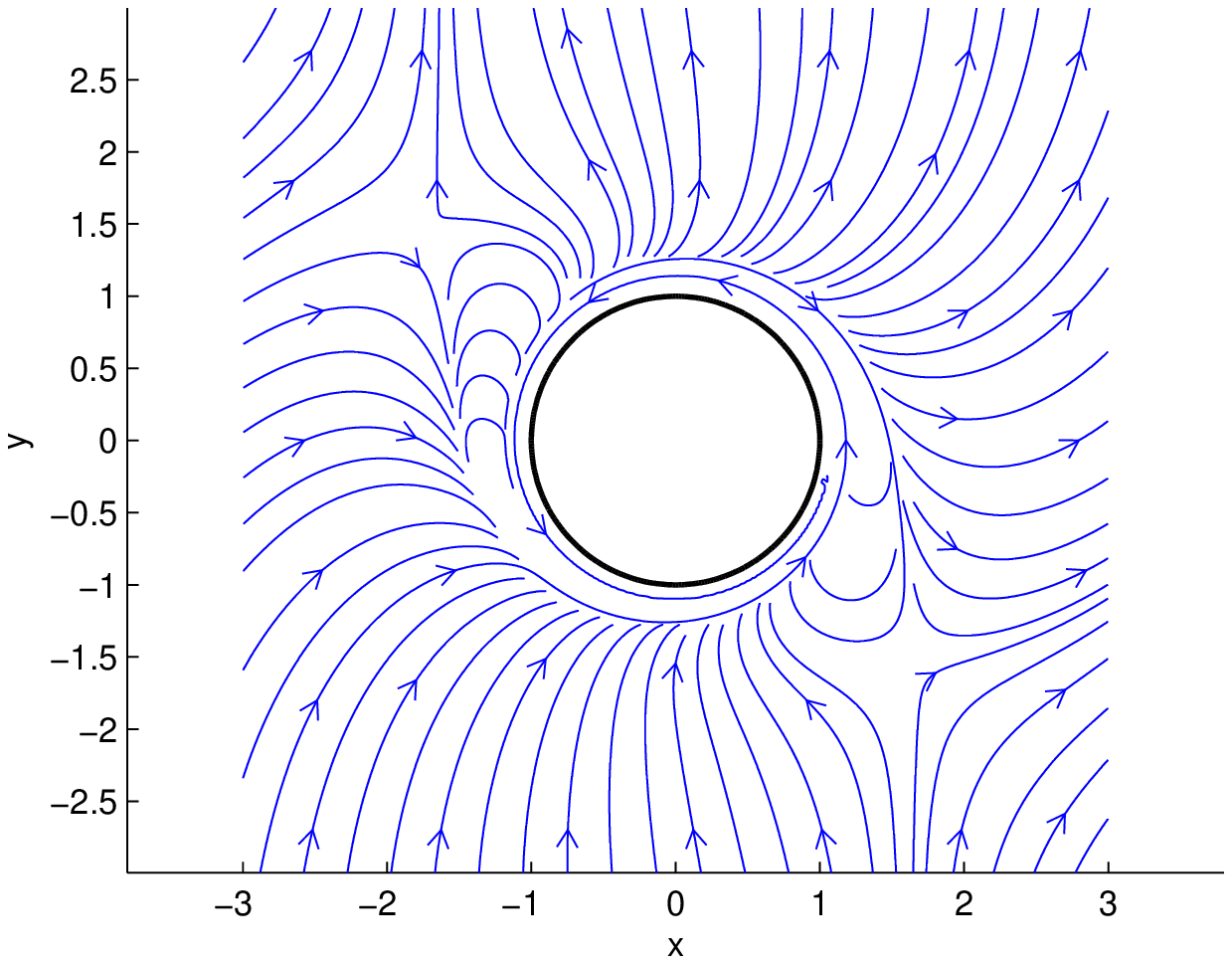}
\caption{Magnetic field lines lying in the equatorial plane $xy$, 
perpendicular to the rotation axis of an extremely rotating BH
 (units are scale with of gravitational radius $r_{\rm{}g}$). Left panel: an
asymptotically uniform magnetic field is directed along the $x$-axis at
large radii and plotted with respect to the physical frame of an
orbiting star. The case of a  co-rotating frame is shown ($a=1$). 
Right panel: example of the projection  of the rotation induced
electric field, corresponding to the magnetic field in the left panel.
\label{fig1}}
\end{figure*}

\begin{figure*}[tbh]
\includegraphics[width=.49\textwidth]{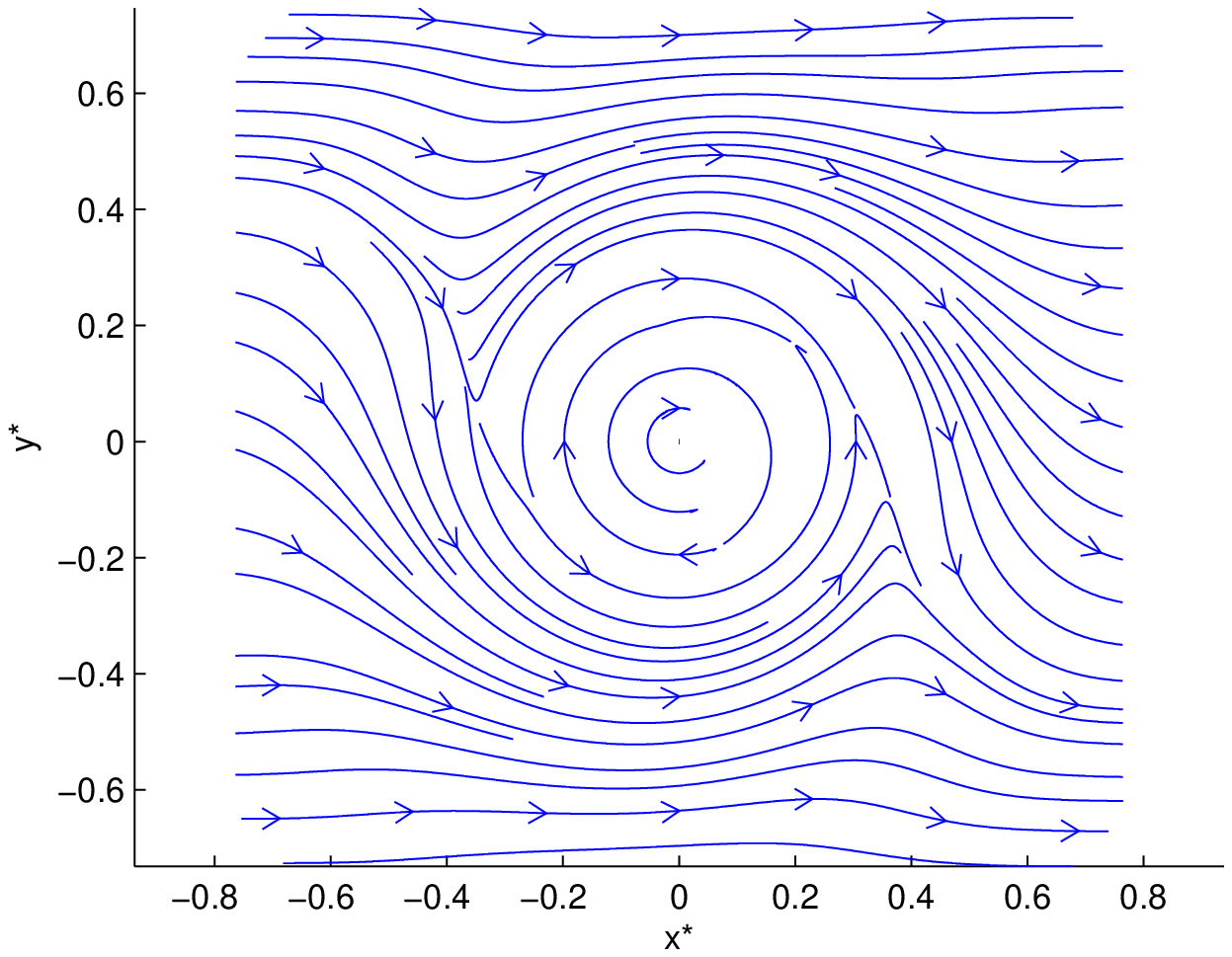}
\hfill
\includegraphics[width=.49\textwidth]{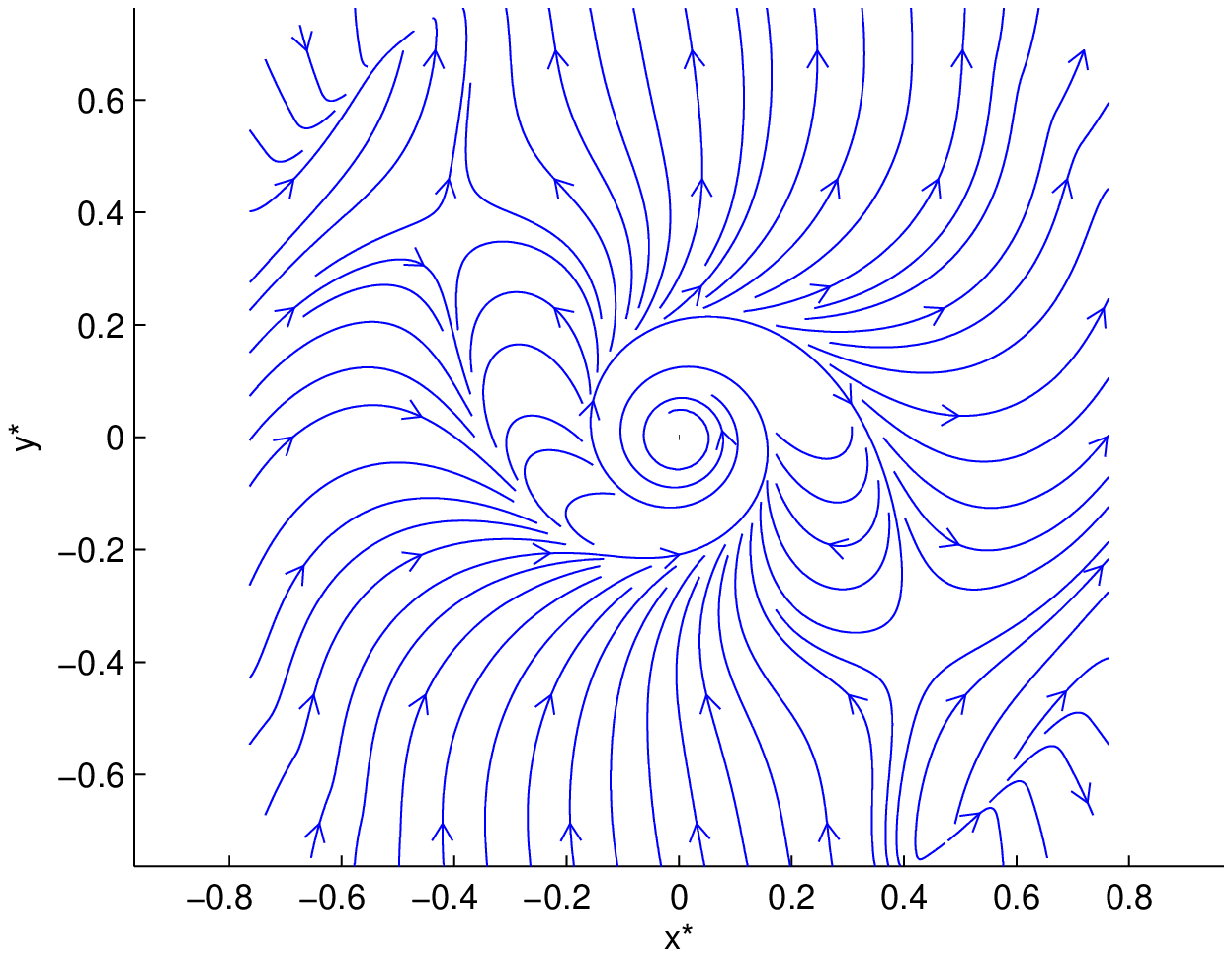}
\caption{The 
field structure as in the left panel of Fig.~\ref{fig1}, now using the
new radial coordinate $r^*$. The magnetic field (left panel) and the
corresponding electric field (right panel) are shown. 
\label{fig2}}
\end{figure*}

\begin{figure*}[tbh]
\includegraphics[width=.39\textwidth]{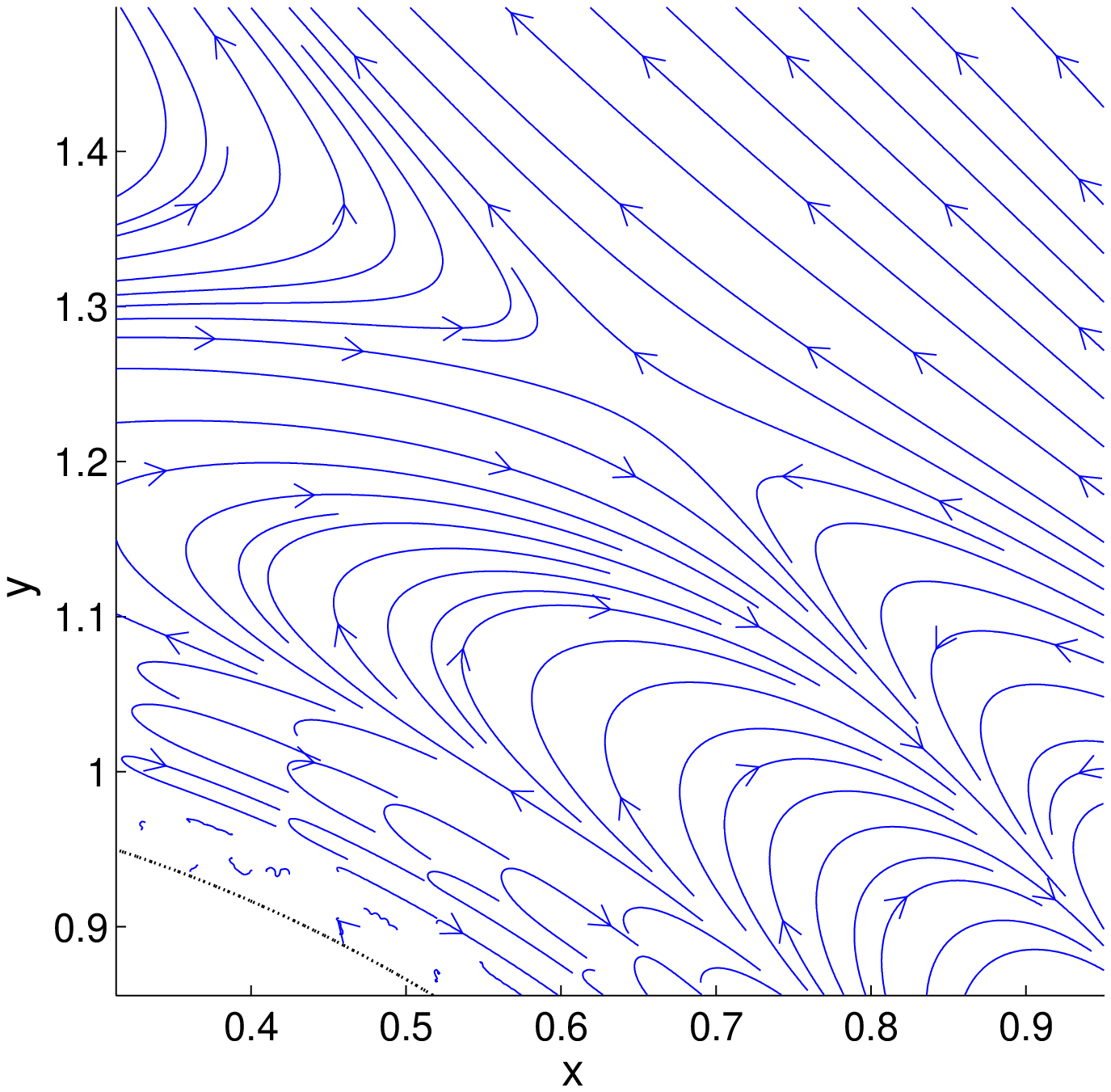}
\hfill
\includegraphics[width=.49\textwidth]{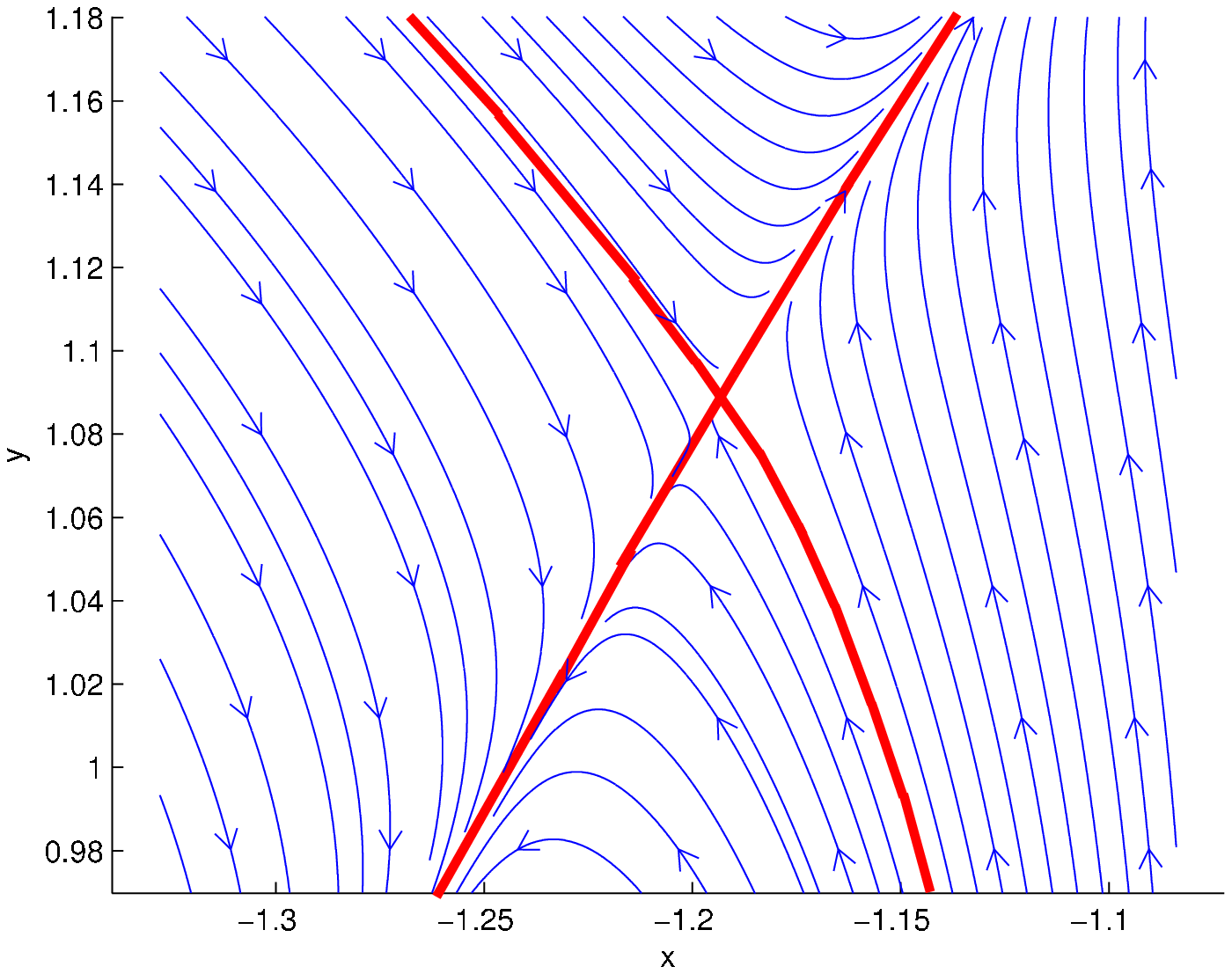}
\caption{Two examples of detailed magnetic structure surrounding a null
point, where the magnetic field vanishes in the black hole equatorial
plane. The separatrix line is shown (solid line by red colour) in the right panel.
\label{fig3}}
\end{figure*}

Hereafter we use spheroidal coordinates with
$\mu\equiv\cos\theta$, $\sigma\equiv\sin\theta$,
and the metric line element in the form
%\begin{eqnarray}
${\rm d}s^{2} = 
  -\Delta\Sigma{A^{-1}}\,{\rd}t^{2}
   +\Sigma\Delta^{-1}\,{\rd}r^{2}
   +\Sigma\,{\rd}\theta^{2}
+A\Sigma^{-1}\sigma^2\;
   \left({\rd}\phi-\omega\,{\rd}t\right)^{2},$
%\label{metric}
%\end{eqnarray}
where $a$ denotes the specific angular momentum, $|a|\leq1$ ($a=0$ is
for a non-rotating BH, while $a=\pm1$ is for the maximally
co/counter-rotating one), $\Delta=r^{2}-2r+a^{2}$,
$\Sigma=r^{2}+a^{2}\mu^2$, $A=(r^{2}+a^{2})^{2}-{\Delta}a^{2}\sigma^2$,
and $\omega=2ar/A$. The outer horizon, $r\equiv r_+(a)$, is where
$\Delta=0$, whereas the ISCO ranges between $r_+(a)\leq
r_{\rm{}isco}(a)\leq6$. The presence of terms
$\propto\omega$ in Kerr metric indicates that frame
dragging operates and affects the motion of particles and the structure of
fields \citep{kv94,mvs99}.

The influence of general relativistic frame-dragging on accreted
particles resembles the effect of a rotating viscous medium: it forces
the particles to share the rotational motion of the central body.
Similarly, it also affects the magnetic field lines and generates the
electric component. For the magnetic field, we might liken the impact of
frame-dragging to the Parker spiral, describing the shape of the
(rotating) Sun's magnetic field. In a way, the action of the
frame-dragging is reminiscent of neutral points
arising due to the interaction of the interplanetary magnetic field with
the Earth dipole \citep{dun61}, but here the strong gravity plays a
major role. Besides other contrasts, the influence of the rotating BH 
grows overwhelmingly as one enters the ergospheric region below
$r=2r_{\rm{}g}$.

The models of electromagnetic acceleration and
collimation of jets have been greatly advanced during the last decade
\citep{fg01,k04,kbv07,khh05}. These works demonstrate the astrophysical
importance of BH rotation, but their setup is somewhat different
from this paper. In particular, the field is typically assumed to be frozen
into an equatorial accretion disc; the magnetic field lines
are twisted by differential rotation of the disc plasma. On the other hand, 
the case of Sgr~A* is distinct, with only a tiny accretion rate,
$\dot{M}\leq10^{-7}M_{\odot}\;{\rm{}yr}^{-1}$ \citep{mf03}, although
any firm estimates are currently not possible because of uncertainties
in the accretion efficiency \citep{ess05}. The standard disc is
not present here; the gas chunks arrive episodically \citep{cm97,cnm08}.
In these circumstances one cannot expect the BH and the ordered
magnetic field to have a common symmetry axis (the Bardeen-Petterson
effect does not operate due to the lack of a steady accretion flow)
neither that the black hole is resting in the centre.

We assume that the electromagnetic field does not contribute to the 
system gravity, which is correct for every astrophysically 
realistic situation.  Within a limited volume around the BH,
typically of size $\simeq(10r_{\rm{}g})^3$, the magnetic field lines
have a structure resembling the asymptotically uniform field. The
electromagnetic field is a potential one and can be written as a
superposition of two parts: the aligned component \citep{kkl75,wal74},
plus the asymptotically perpendicular field \citep{bd76,bj80}.
The four-potential has been given explicitly in terms of functions
$\psi\equiv\phi+a\delta^{-1}\ln\left[\left(r-r_+\right)/\left(r-r_-\right)\right]$
is the Kerr ingoing angular coordinate, $\Psi=r\cos\psi-a\sin\psi$,
$\delta=r_+-r_-$, and $r_{\pm}=1\pm\sqrt{1-a^2}$ \citep{bj80,k08}. 
It is exactly the variable $\psi$ that determines the growing twist
of magnetic lines, which eventually leads to the formation of
magnetic null points near a rotating black hole.
We can thus employ these expressions to draw lines of force.

A particle can be accelerated by the equipartition field, acting along 
the distance $\ell$, to energy $\gamma_{\rm{}max}\simeq
10^{18}q_{\rm{}e}(B/10G)\,(\ell/r_{\rm{}g})$~eV, where $q_{\rm{}e}$
is in units of the elementary charge. Naturally, this rough estimate can
be exceeded if a non-stationary field governs the acceleration
process.

The aligned vacuum field is gradually expelled out of the BH as its
rotation increases \citep{kkl75}. Non-vacuum fields are more complicated
and have been studied in detail; e.g.\ \citet{km07} notice that
high conductivity of the medium changes the field expulsion properties
of the horizon. On the other hand, only few aspects of the misaligned BH
magnetic fields have been explored so far \citep{na07,t00}. This is because of an extra complexity 
caused by the frame-dragging \citep{ag89,k89}. What we put forward here is that 
the frame-dragging can actually contribute to the acceleration mechanisms.

To obtain the physical components of the electromagnetic tensor,
$F=2dA$, we project it onto the local observer tetrad,
$\bm{e}_{(a)}$ \citep{mtw73}. The appropriate choice of the projection tetrad is the
one attached to a frame in Keplerian orbital motion, which exists for
$r\geq r_{\rm{}isco}$. Free circular motion can be stable at
$r\geq r_{\rm{}isco}$, otherwise the star has to spiral downwards while
maintaining constant angular momentum of $l=l(r_{\rm{}isco})$. The
corresponding frame and the field lines can be derived in a
straightforward manner \citep{k08}. We interpret the frame, quite naturally, as 
the one connected with an orbiting star. In the extremely rotating case, 
the circular motion of a star is
possible all the way down to $r_{\rm{}isco}=r_{\rm{}g}$.
The dependence of the electromagnetic components on the $\psi$ angle
then indicates the ever increasing effect of the frame dragging near the
horizon.

The electric and magnetic intensities, measured by the physical
observer, are:  $E_{(a)}=\bm{e}_{(a)}^{\mu}F_{\mu\nu}u^{\nu}$,
$B_{(a)}=\bm{e}_{(a)}^{\mu}F^{*}_{\mu\nu}u^{\nu}$, where
$u^\nu\equiv\bm{e}_{(t)}^{\nu}$ is the observer's four-velocity (the
remaining three basis vectors can be conveniently chosen as space-like,
mutually perpendicular vectors). 

Figure~\ref{fig1} shows the typical structure of the field lines
representing an asymptotically transverse magnetic field
($B_{\parallel}=0$). Formation of the layers in the ergospheric region
just above ISCO is an interesting and generic feature of the rotating
spacetime. Two critical points can be seen occurring at radii up to
$\simeq1.7r_{\rm{}g}$ for $a=1$. Naturally, the co-rotating case ($a>0$)
is not equivalent with the counter-rotating ($a<0$) one, but in both, a
site of oppositely directed field lines arises.

An example of electric lines of force is plotted in the right panel of
Fig.~\ref{fig1}. We notice that the electric field along the magnetic
lines does not vanish and is capable of accelerating charged particles.
Figure~\ref{fig2} then reveals the shape of the near-horizon field lines
in terms of the new radial variable, $r^*\equiv1-r_{\rm{}g}/r$, which we
introduce to better resolve the complicated structure
(Cartesian-type coordinates are used, $x^{*2}+y^{*2}=r^{*2}$). 

So far we have assumed zero translational motion of the BH with
respect to the magnetic field. However, our approach can be generalised
and the uniform motion can be taken into account by Lorentz boost of the
field intensities $E_{(a)}$, $B_{(a)}$.  To this end we notice that the
Lorentz transform, when applied in the asymptotically distant  region
(i)~changes the direction of the uniform magnetic field by an
$\vartheta$, and (ii)~generates a new (uniform) electric component. The
desired form of the electromagnetic test field near a moving black hole
is thus written in a symbolic way,
$F^\prime_{r\rightarrow\infty}=\Lambda(\beta)^{\rm T}
F\Lambda(\beta)$, and obtained as superposition of the two
parts combined together in the right ratio, i.e.\ the asymptotically
uniform magnetic field, rotated into the desired direction, and the
solution for the asymptotically uniform electric field. The latter one
can be found by applying the dual transform to Wald's
field.

\begin{figure*}[ptbh]
\centering
\includegraphics[scale=0.28, clip]{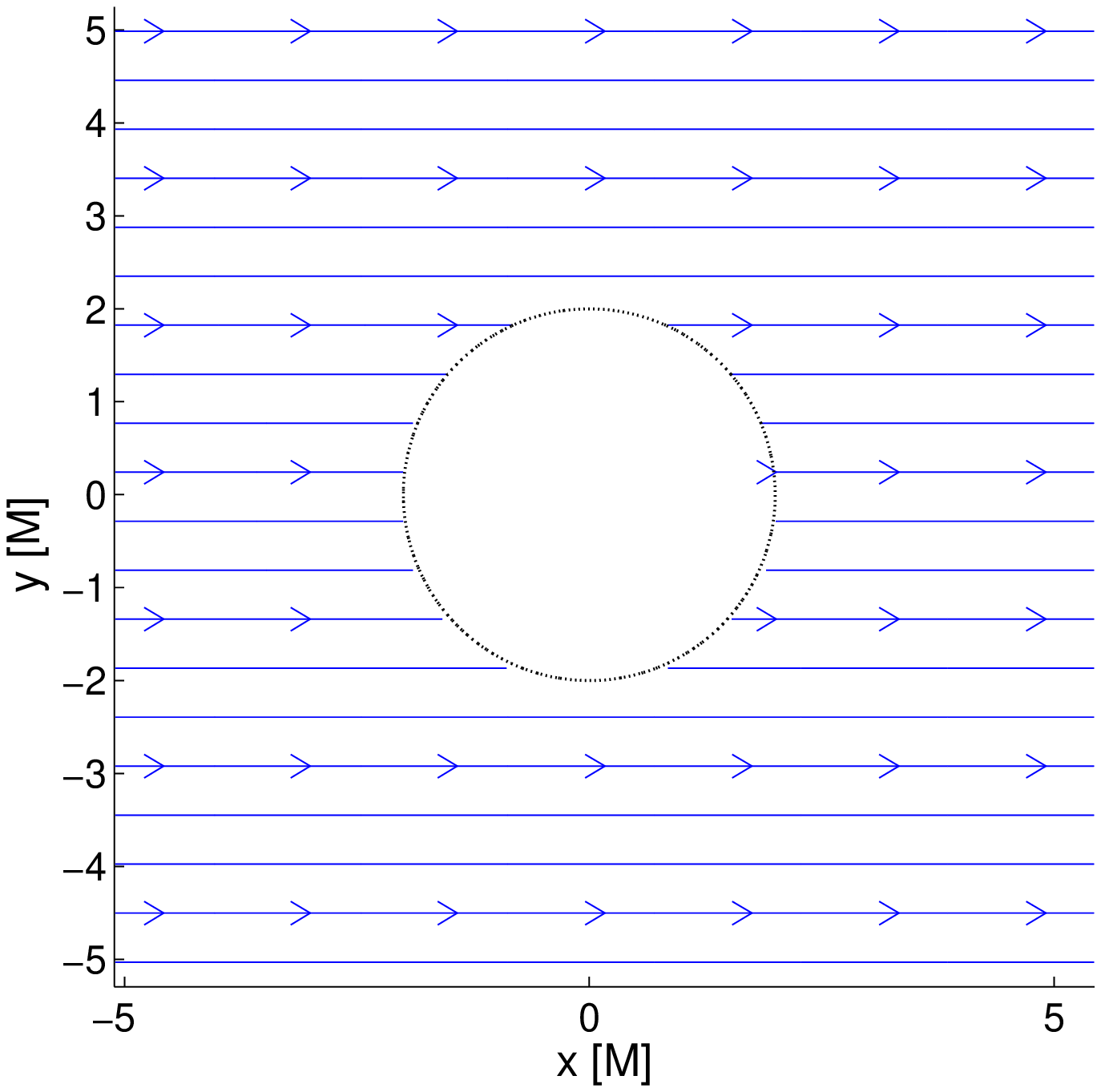}
\includegraphics[scale=0.28, clip]{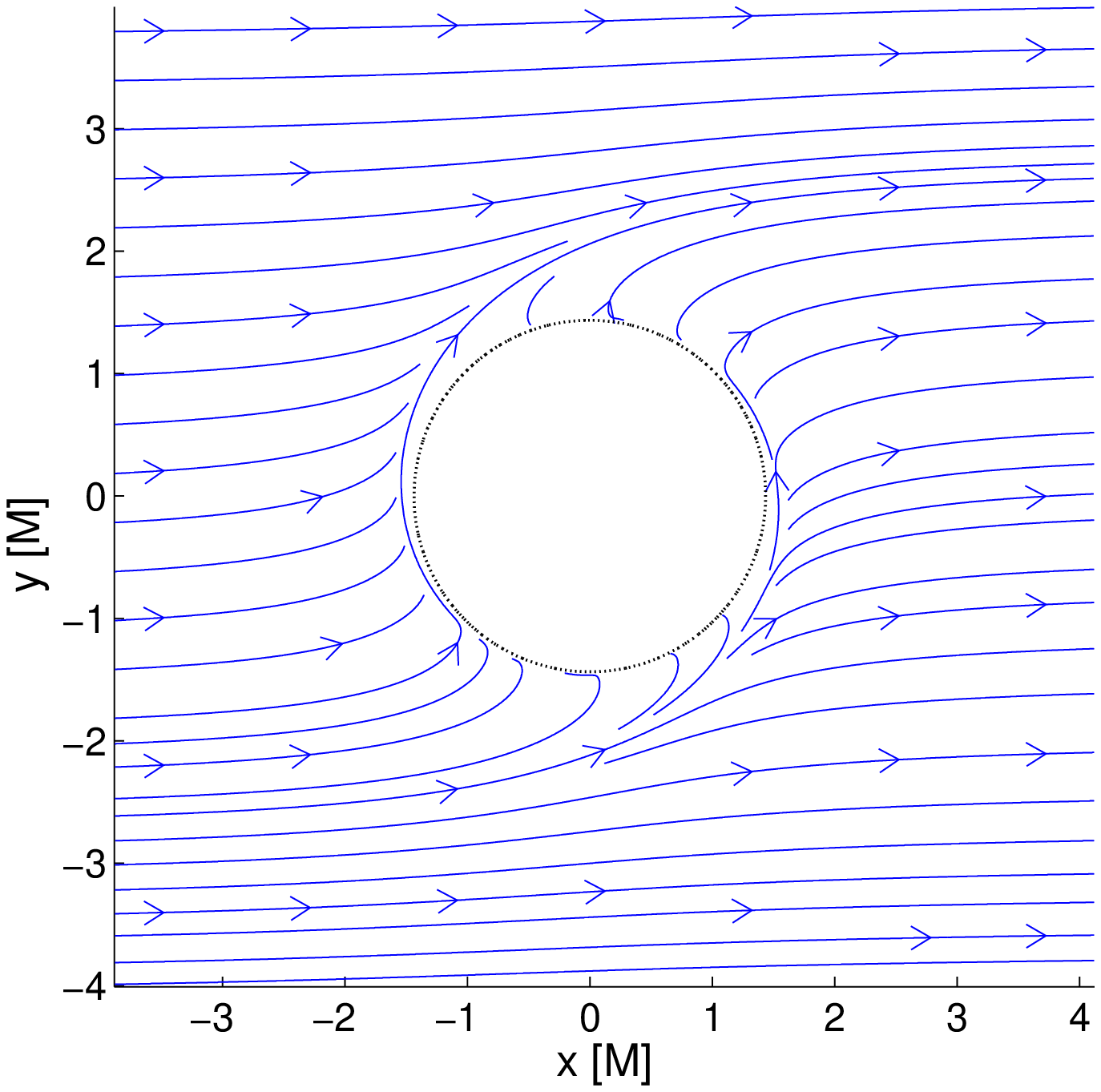}
\includegraphics[scale=0.28, clip]{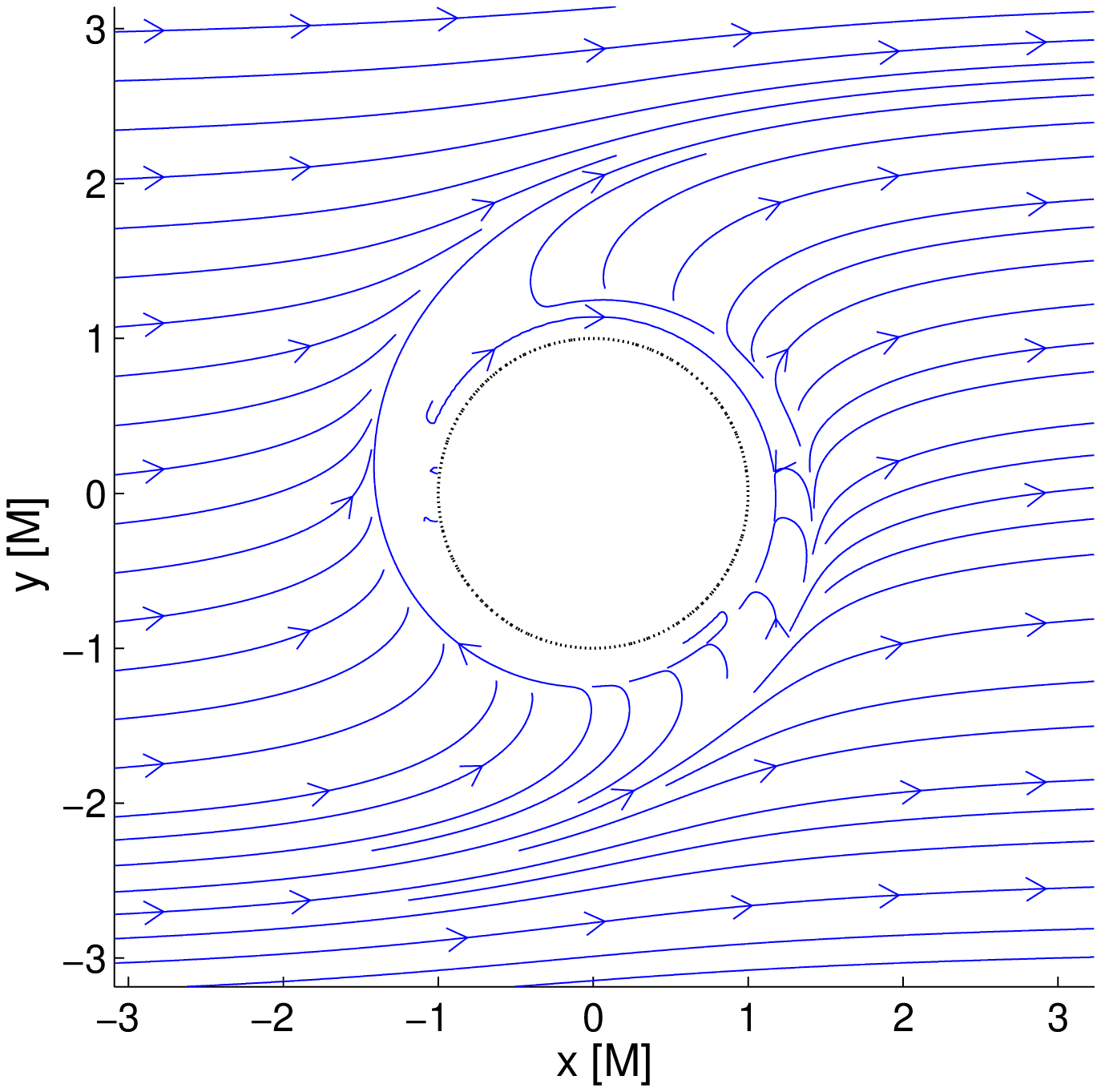}
\includegraphics[scale=0.28, clip]{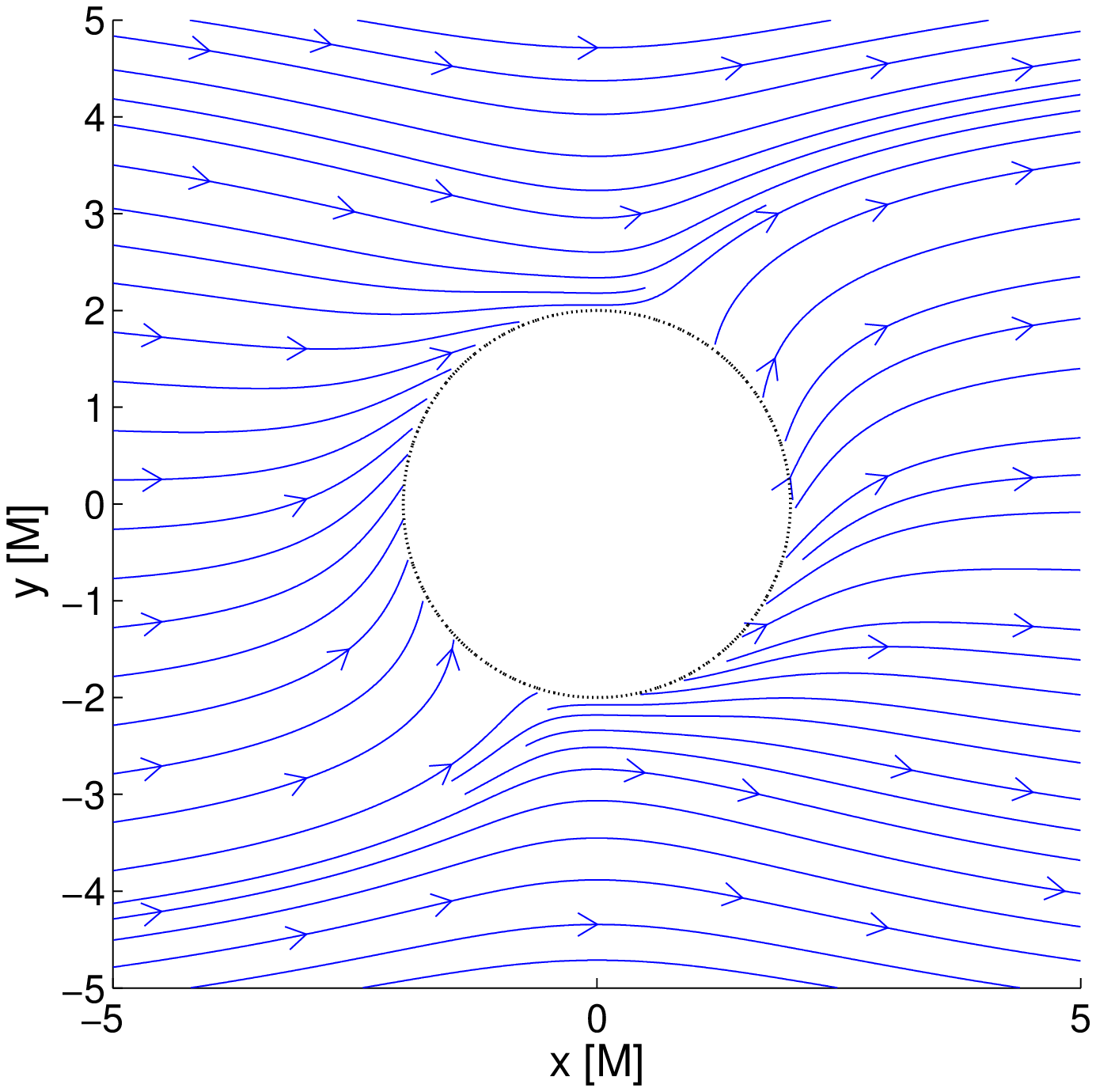}
\includegraphics[scale=0.28, clip]{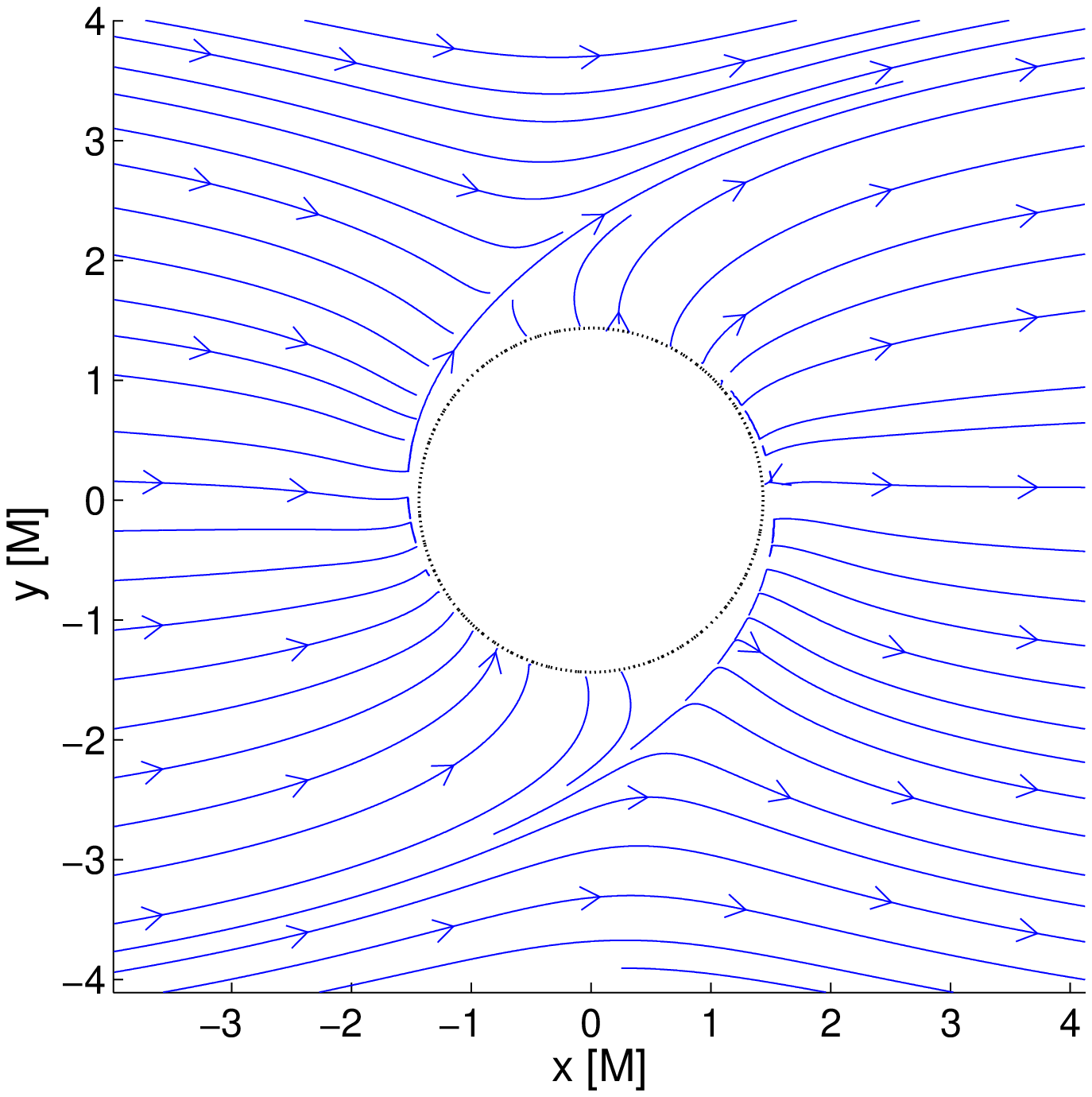}
\includegraphics[scale=0.28, clip]{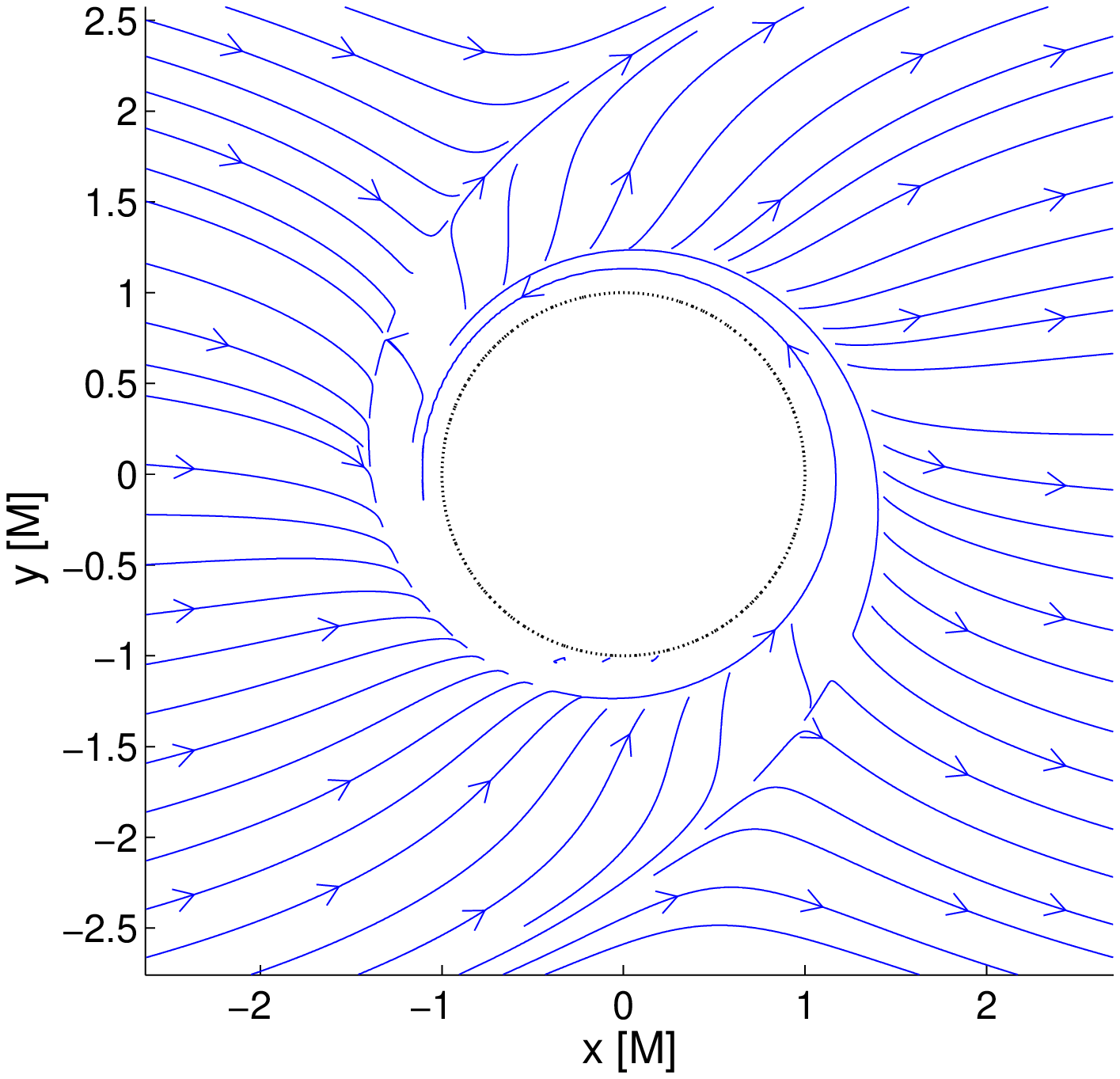}
\includegraphics[scale=0.28, clip]{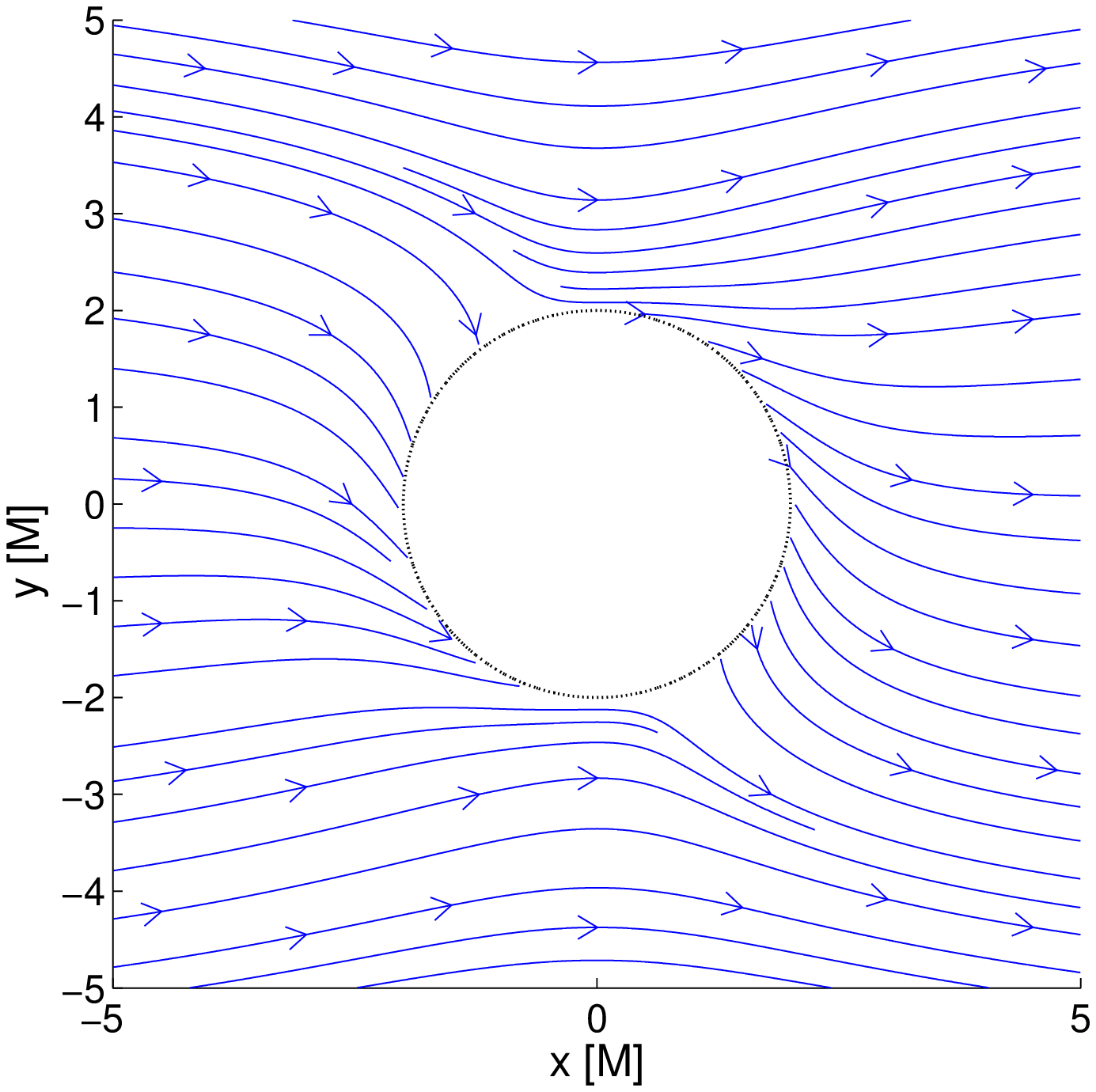}
\includegraphics[scale=0.28, clip]{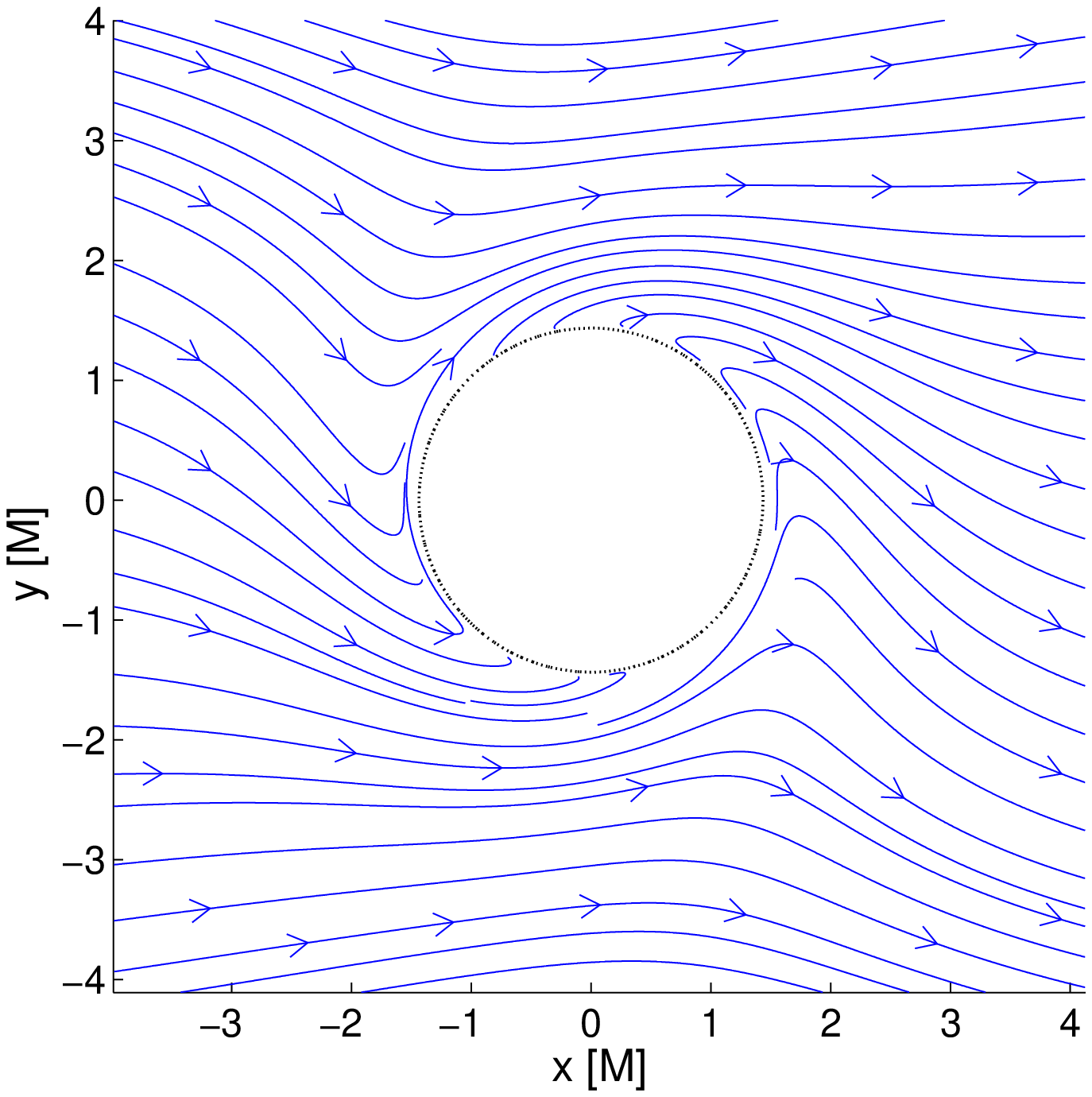}
\includegraphics[scale=0.28, clip]{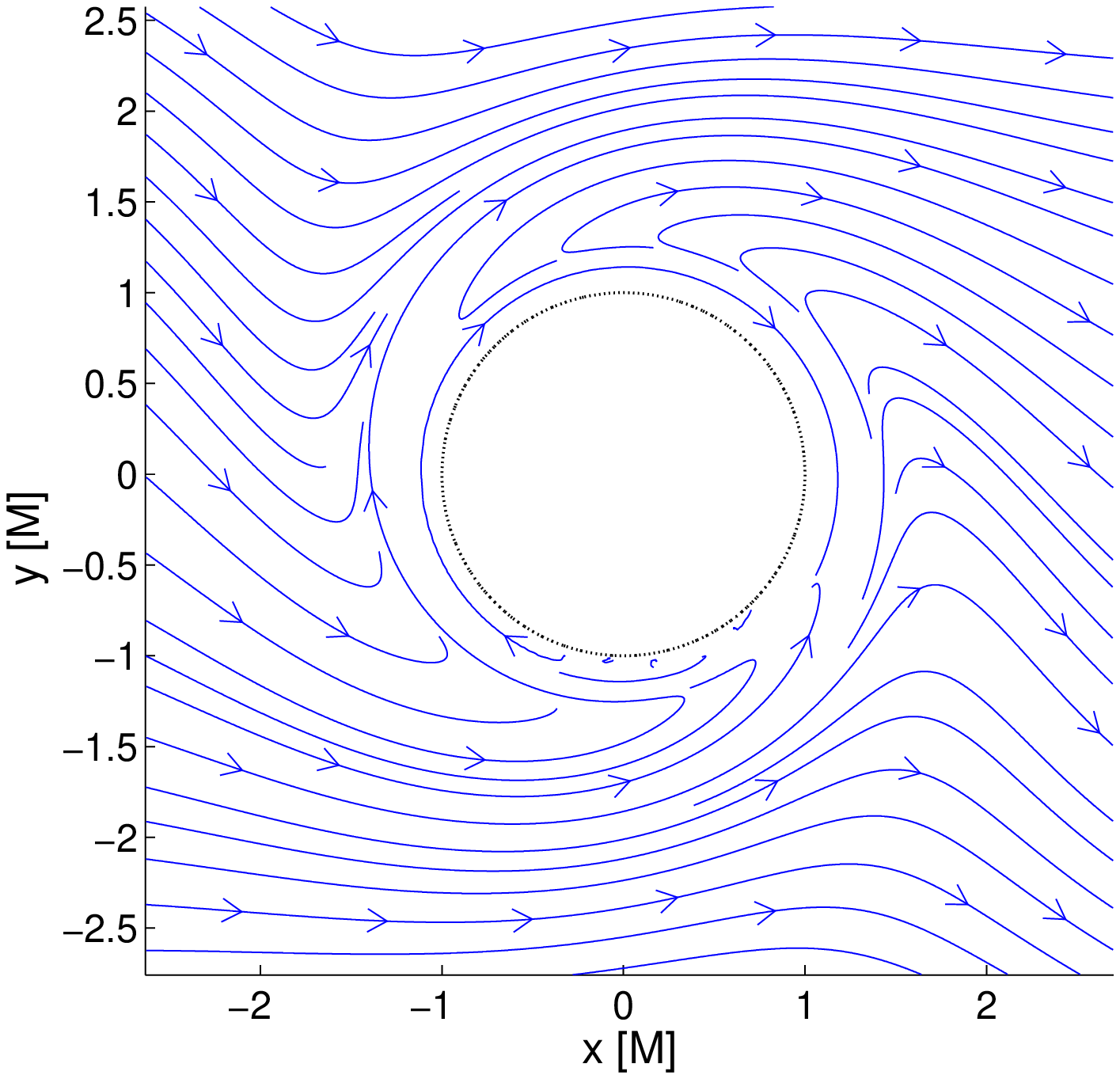}
\includegraphics[scale=0.28, clip]{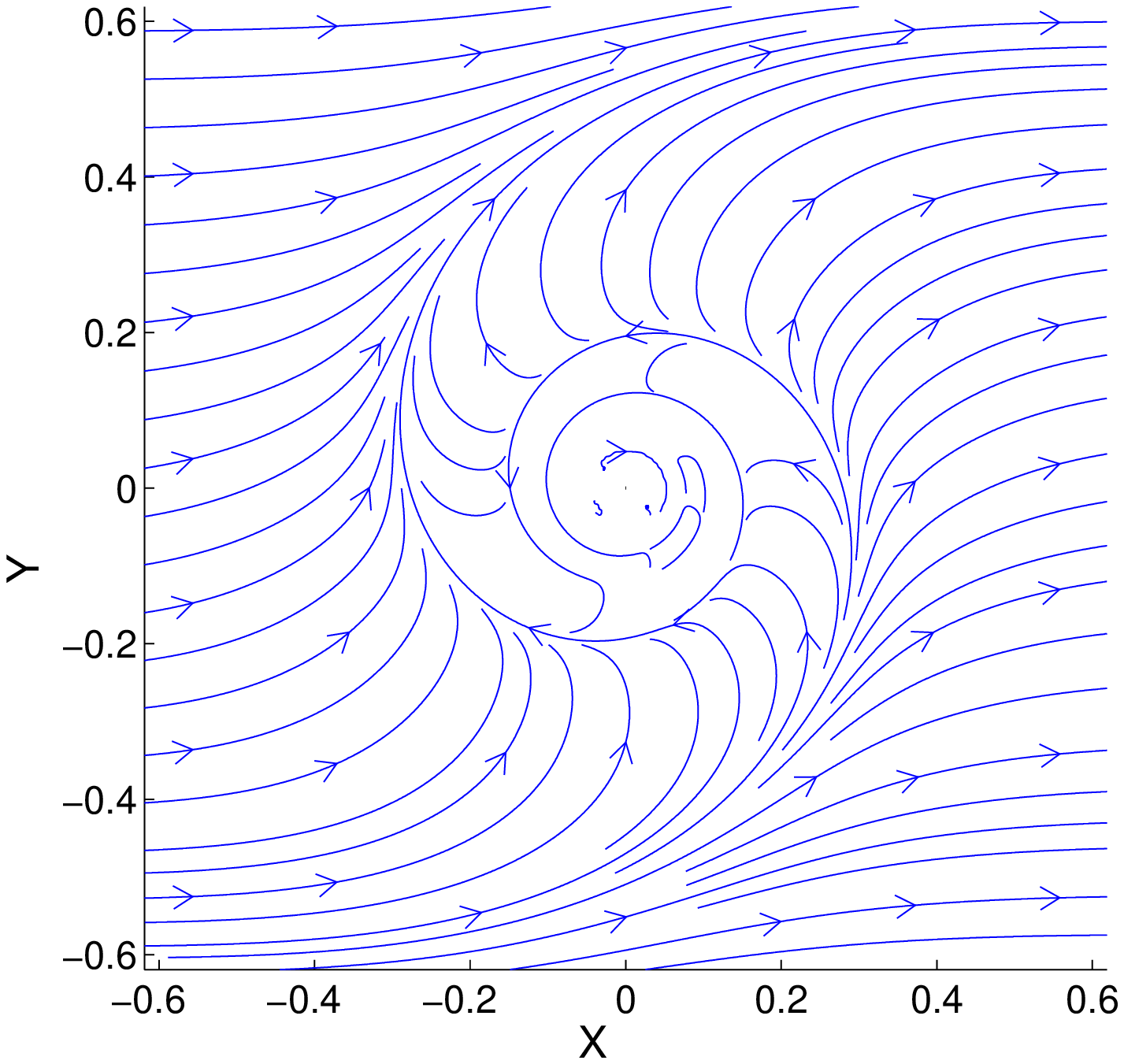}
\includegraphics[scale=0.28, clip]{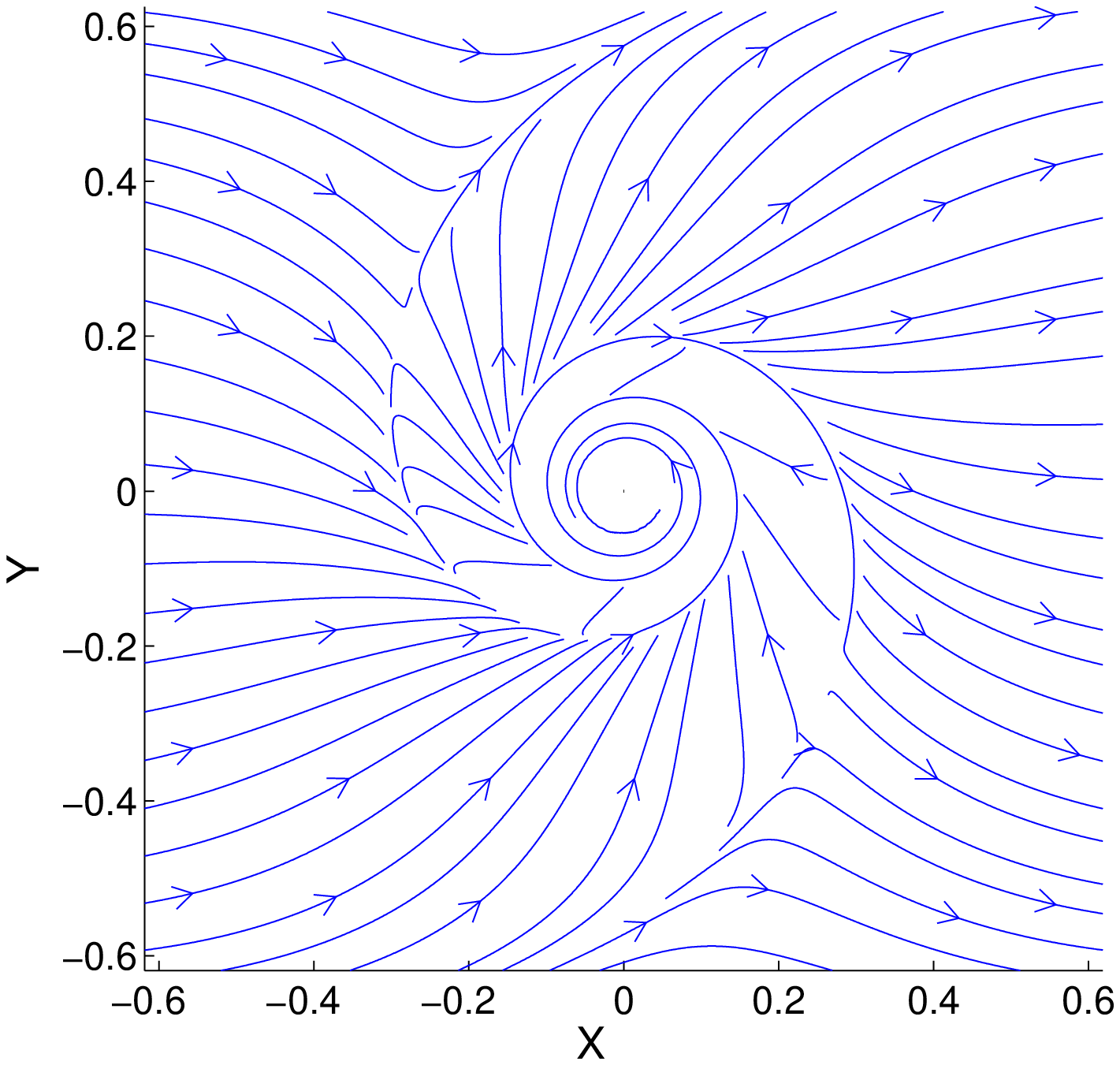}
\includegraphics[scale=0.28, clip]{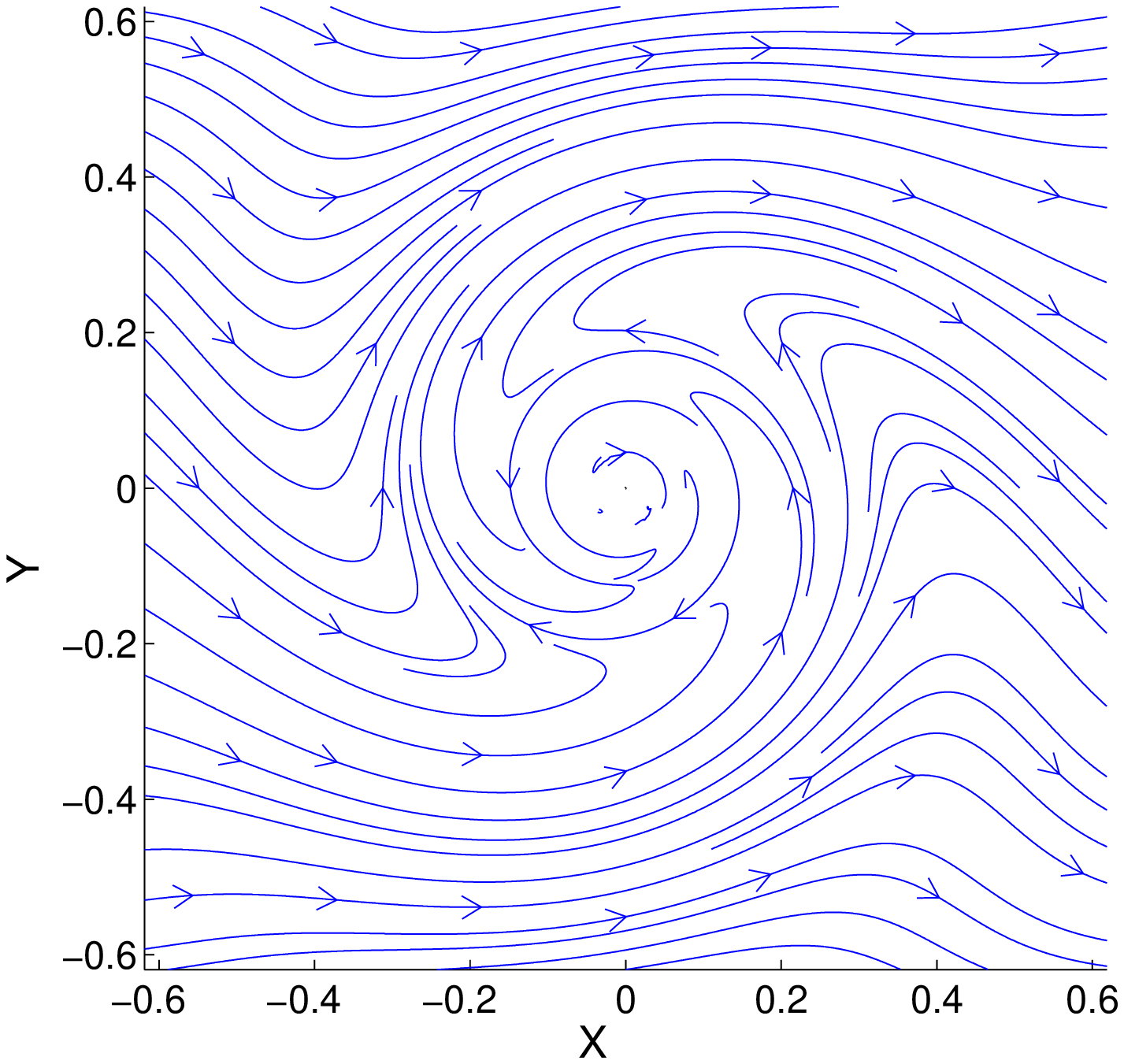}
\caption{Equatorial behaviour of the magnetic field with the purely perpendicular 
magnetic asymptotes 
($B_x\ne 0$, $B_y=B_z=0$) for three distinct tetrads (see the main text for details).}
\label{mag_ekv}
\end{figure*}

\begin{figure}[ptbh]
\centering
\includegraphics[scale=0.295, clip]{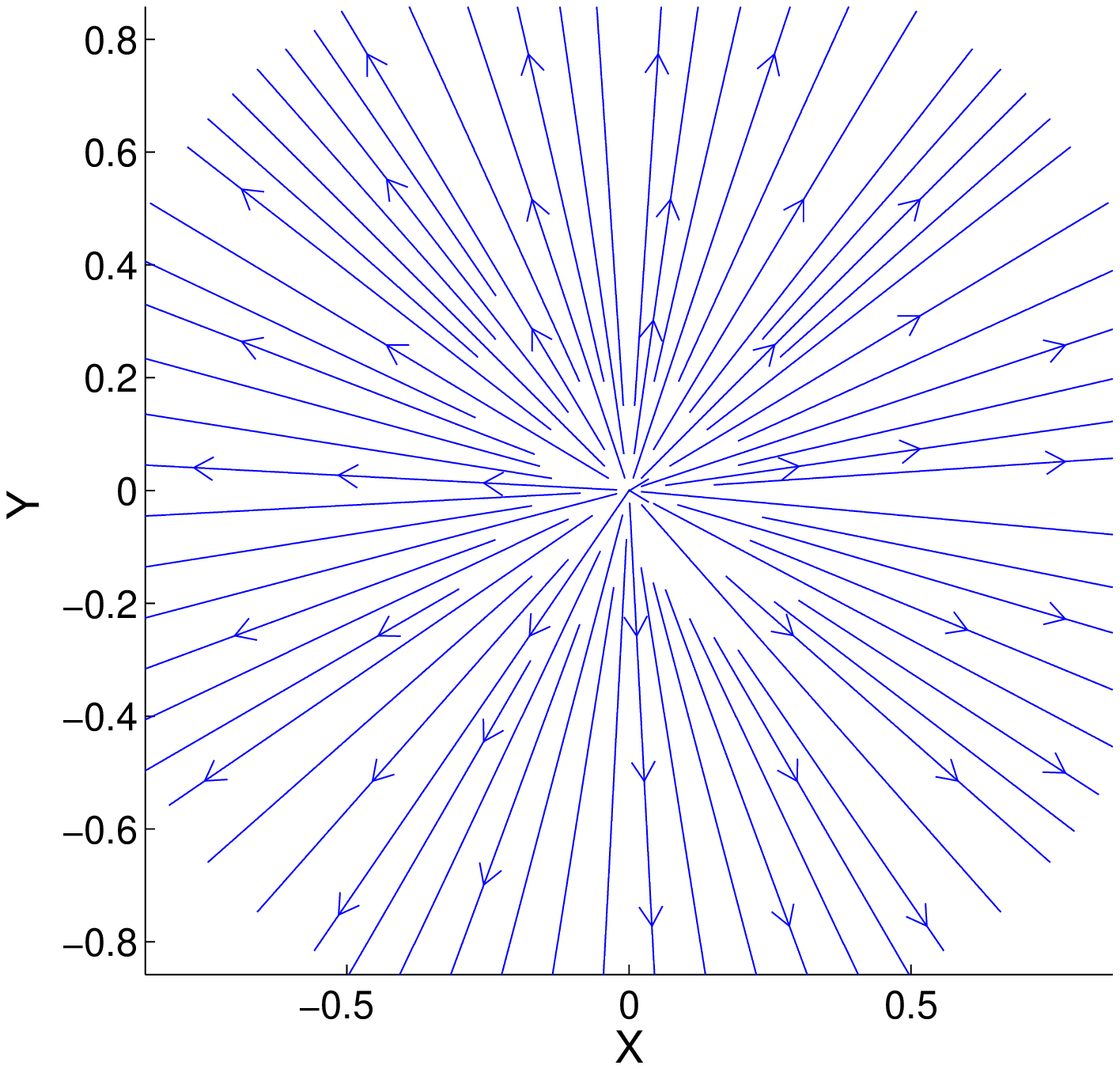}
\includegraphics[scale=0.295, trim= 8mm 0mm 0mm 0mm, clip]{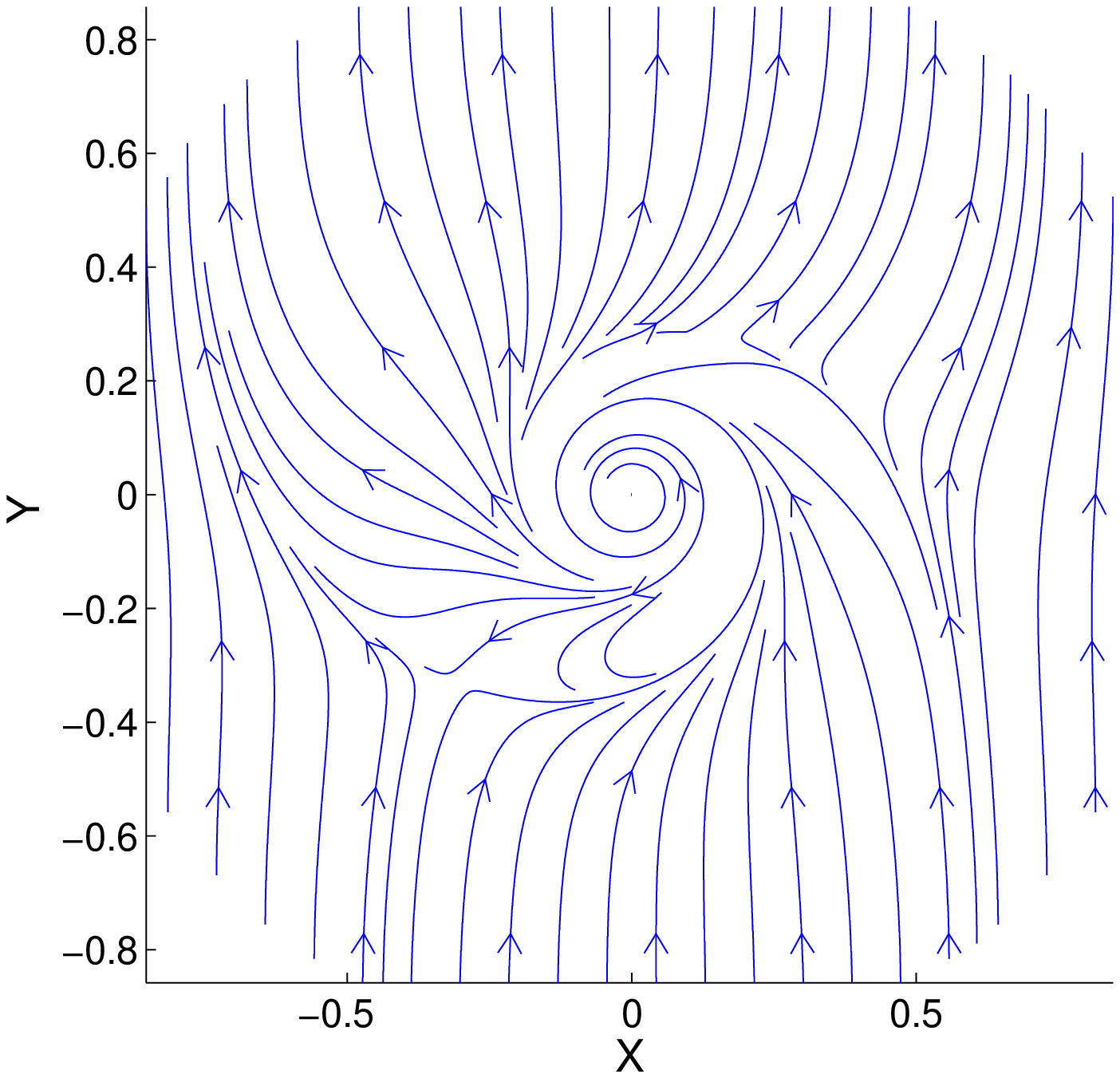}
\includegraphics[scale=0.295, trim= 8mm 0mm 0mm 0mm, clip]{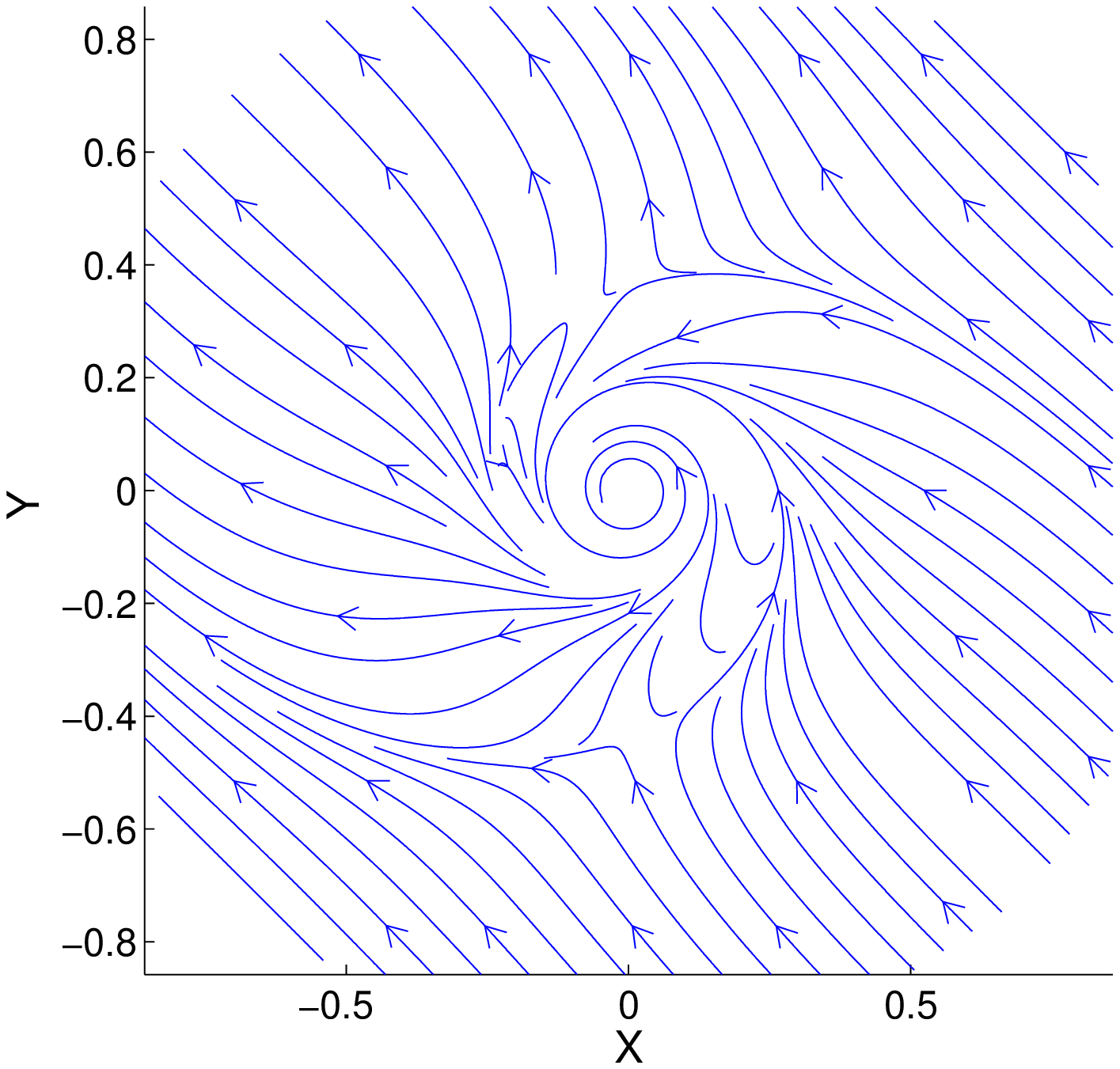}
\includegraphics[scale=0.295, clip]{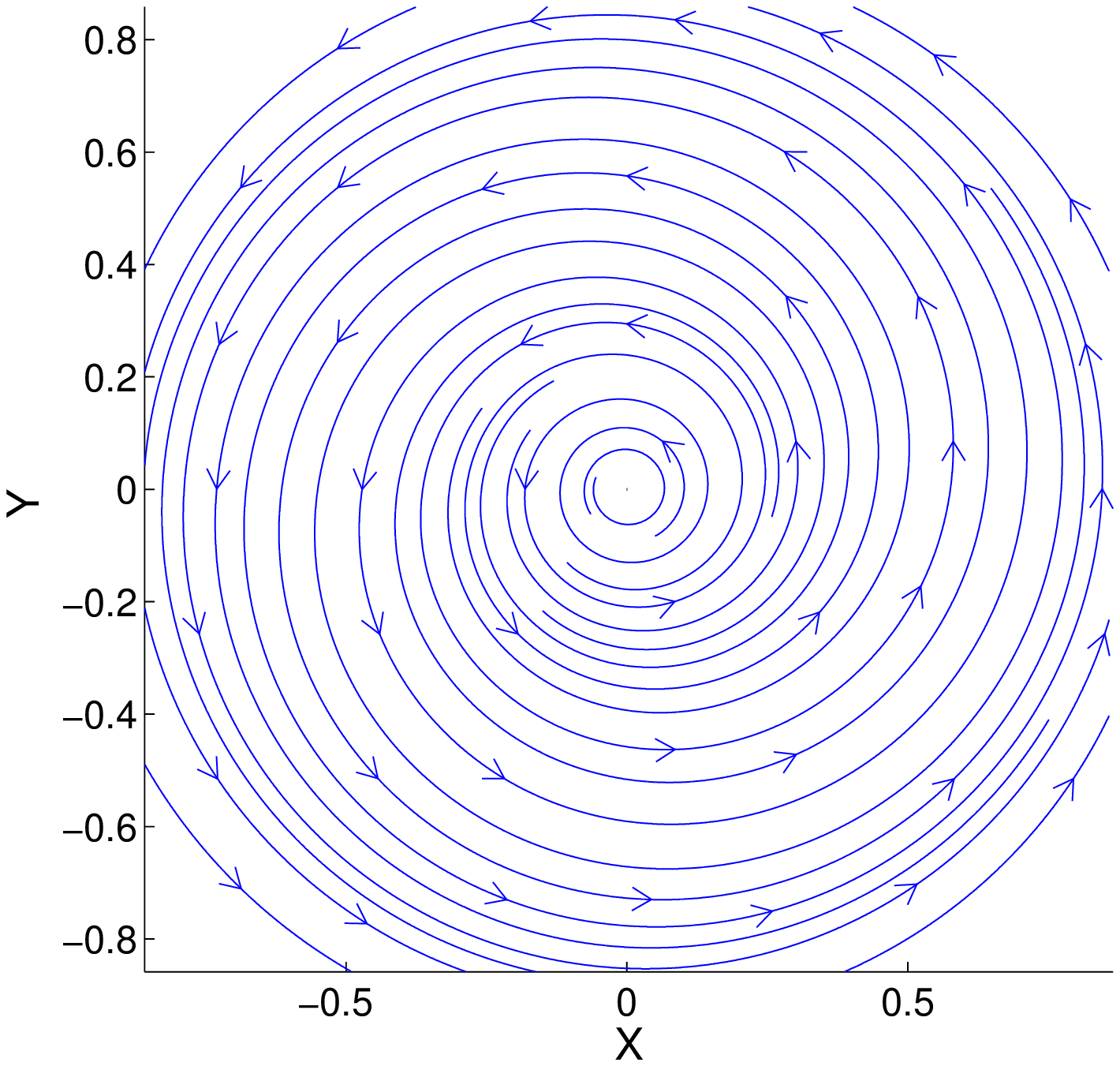}
\includegraphics[scale=0.295, trim= 8mm 0mm 0mm 0mm,clip]{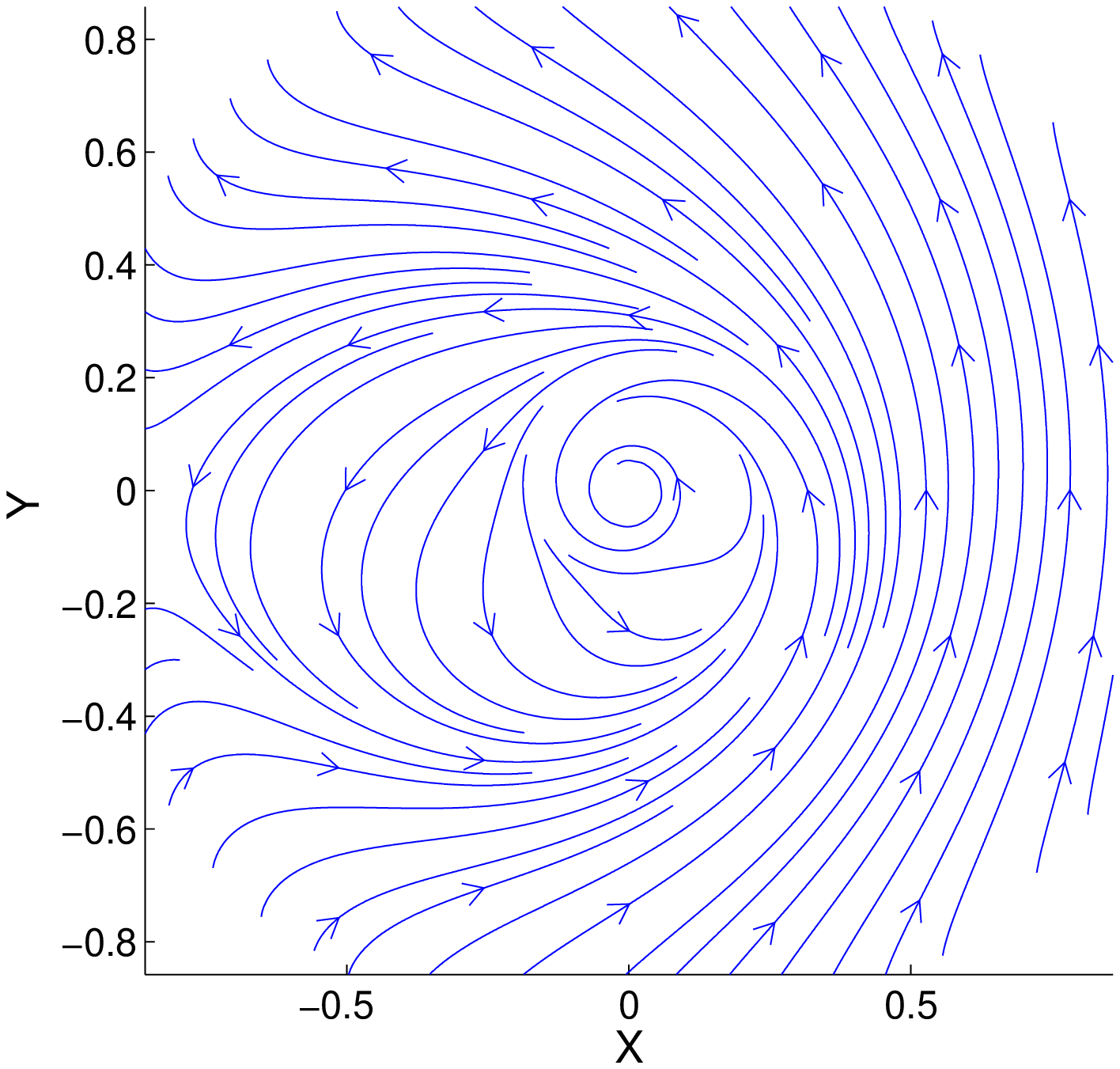}
\includegraphics[scale=0.295, trim= 8mm 0mm 0mm 0mm,clip]{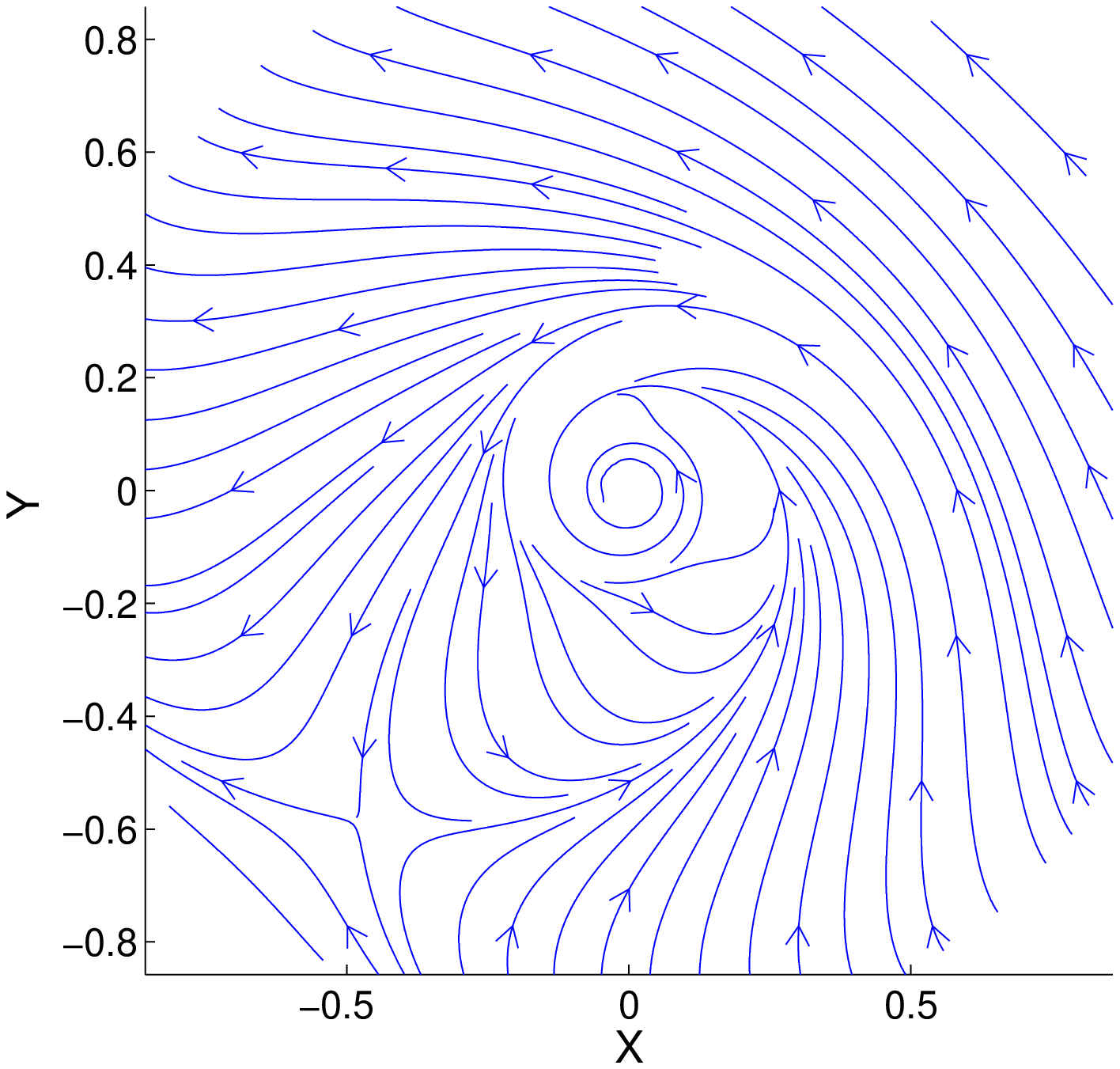}
\includegraphics[scale=0.295, clip]{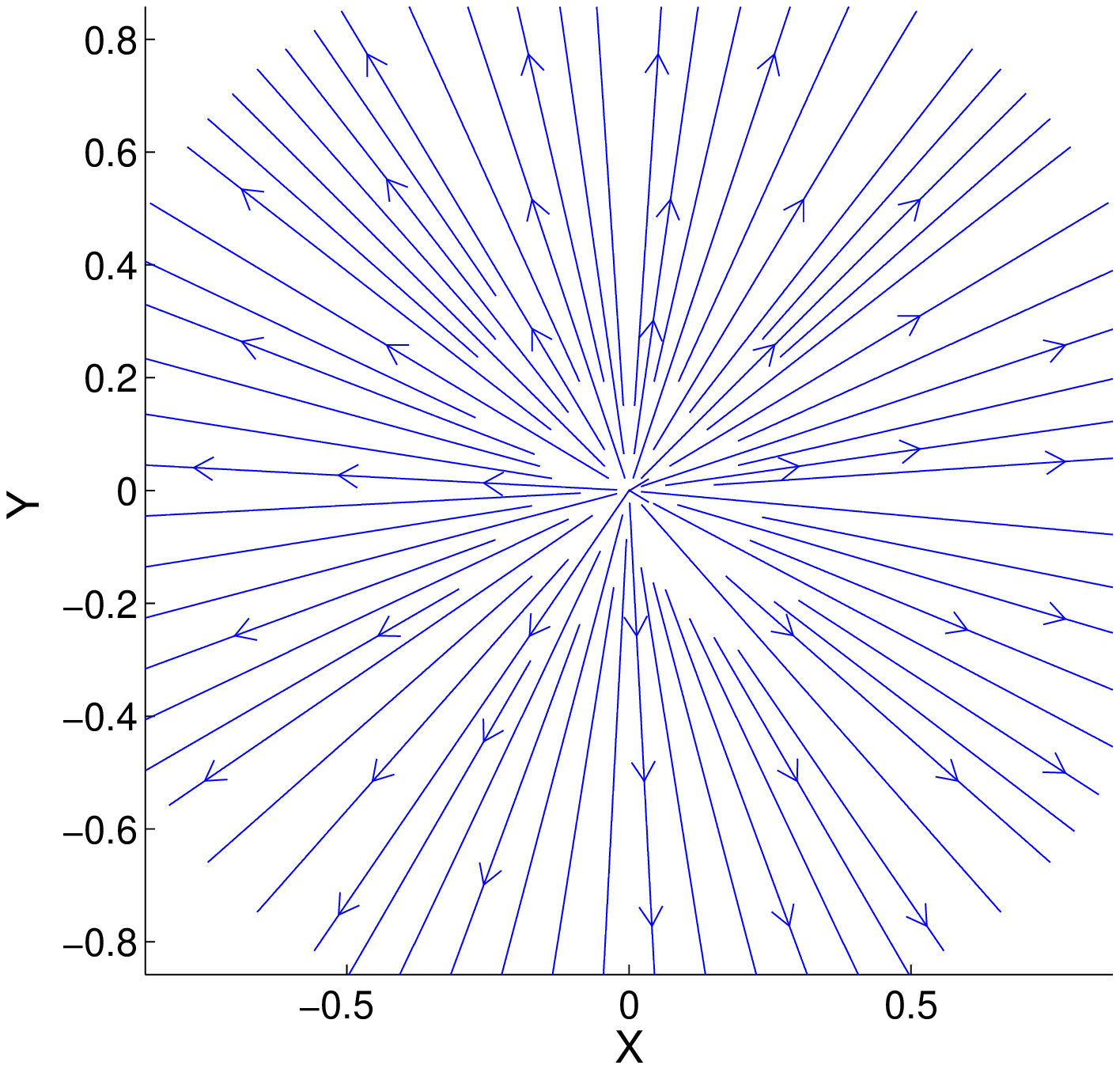}
\includegraphics[scale=0.295, trim= 8mm 0mm 0mm 0mm,clip]{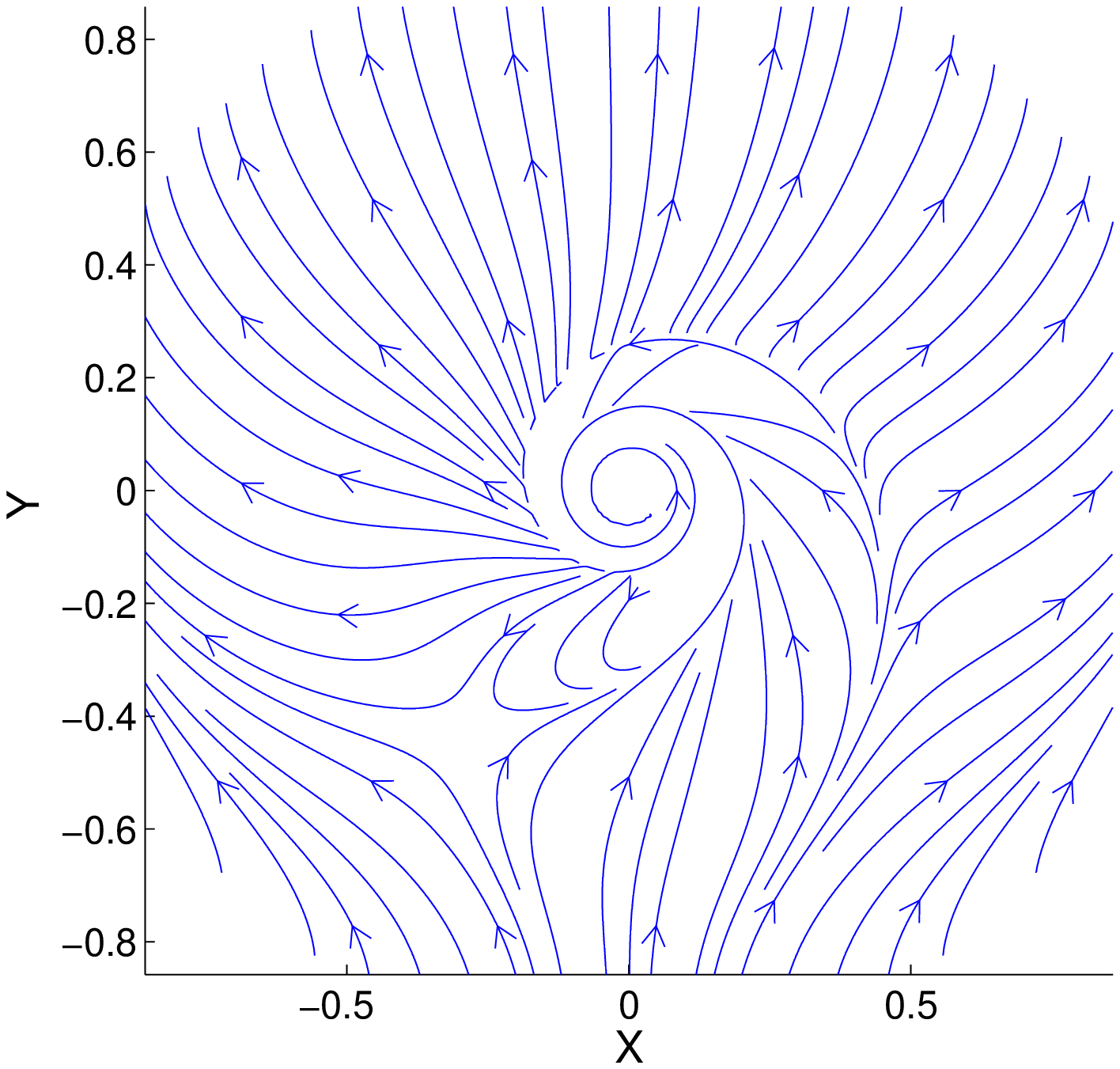}
\includegraphics[scale=0.295, trim= 8mm 0mm 0mm 0mm,clip]{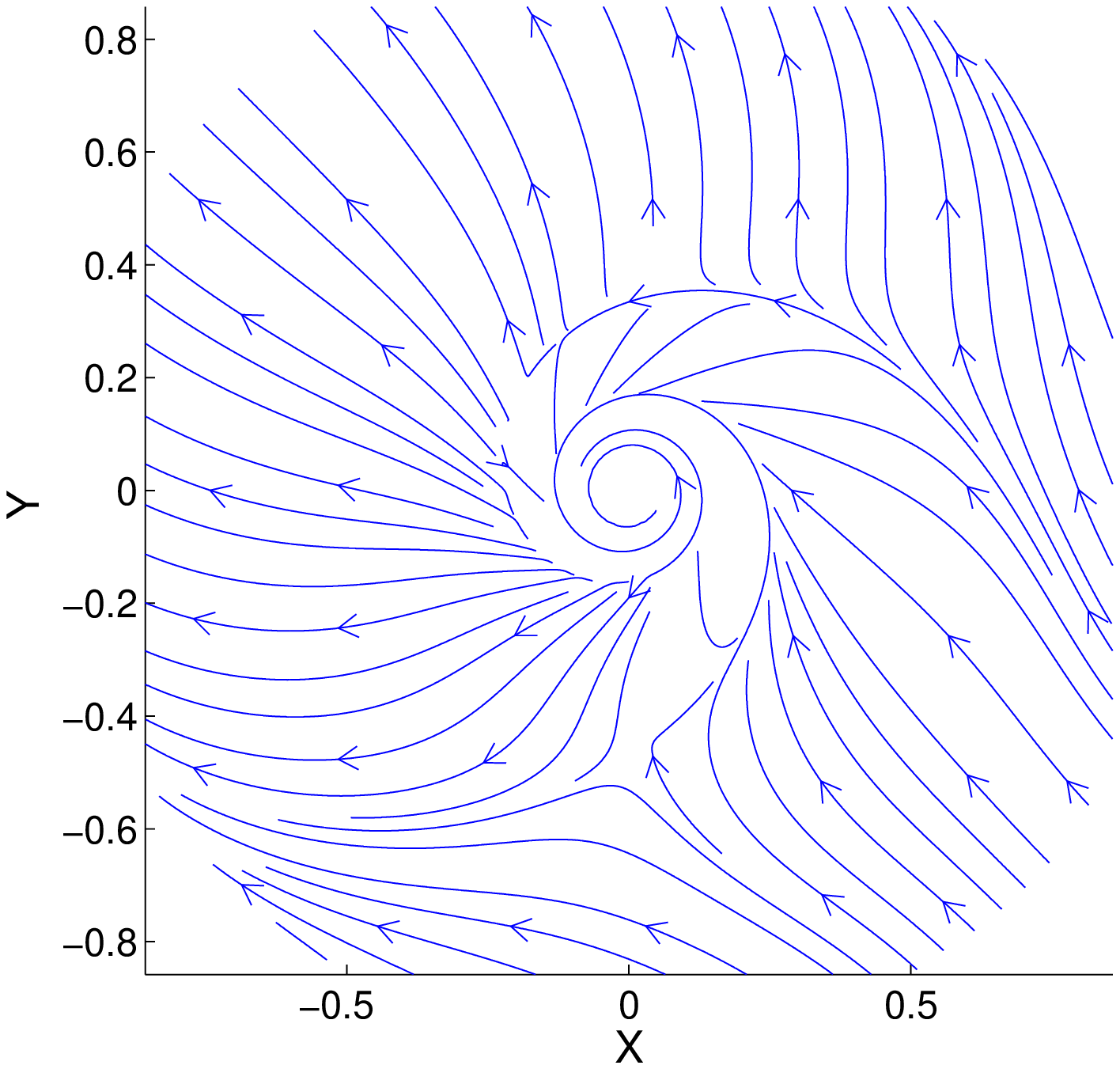}
\includegraphics[scale=0.295, clip]{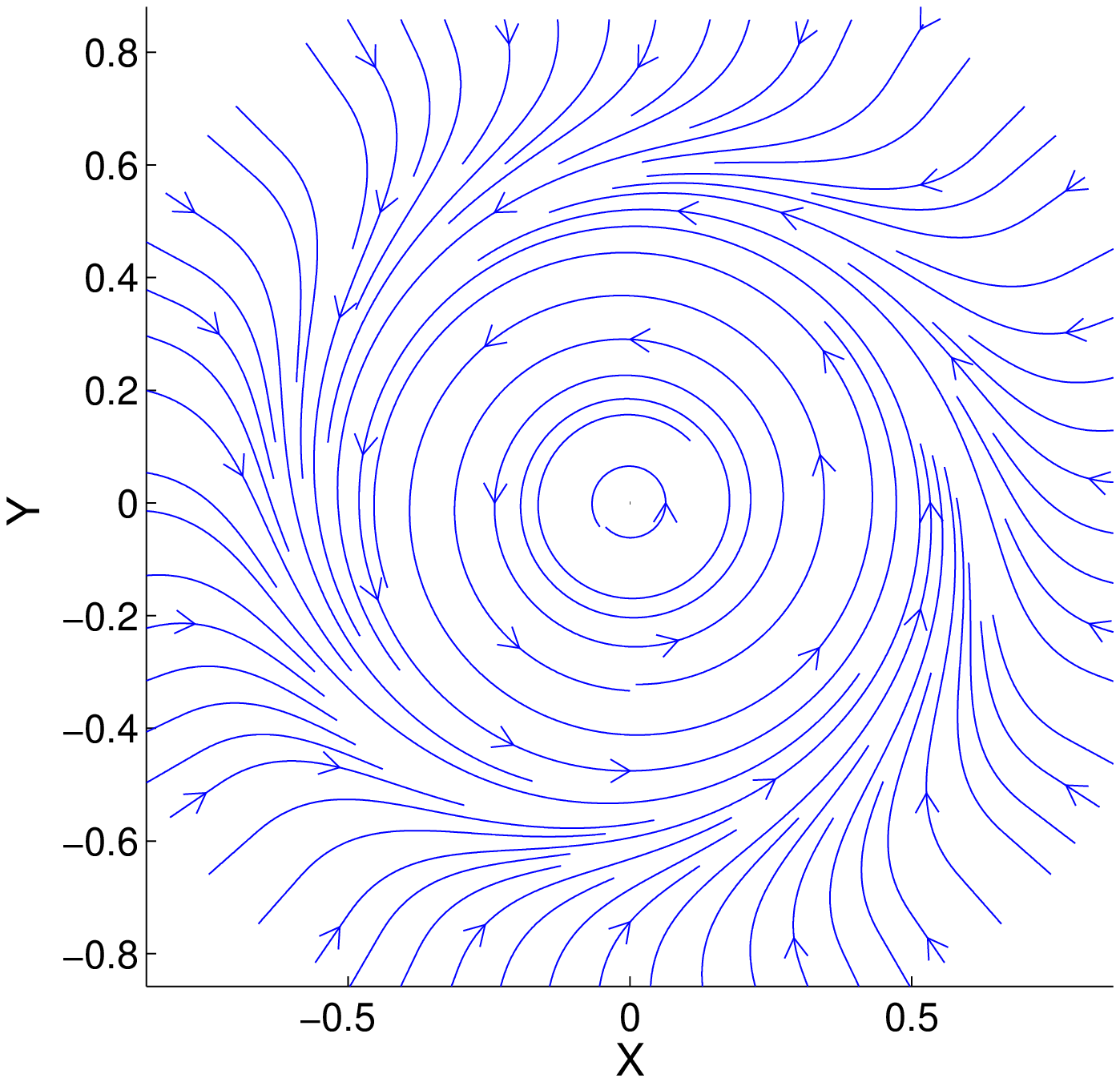}
\includegraphics[scale=0.295, trim= 8mm 0mm 0mm 0mm,clip]{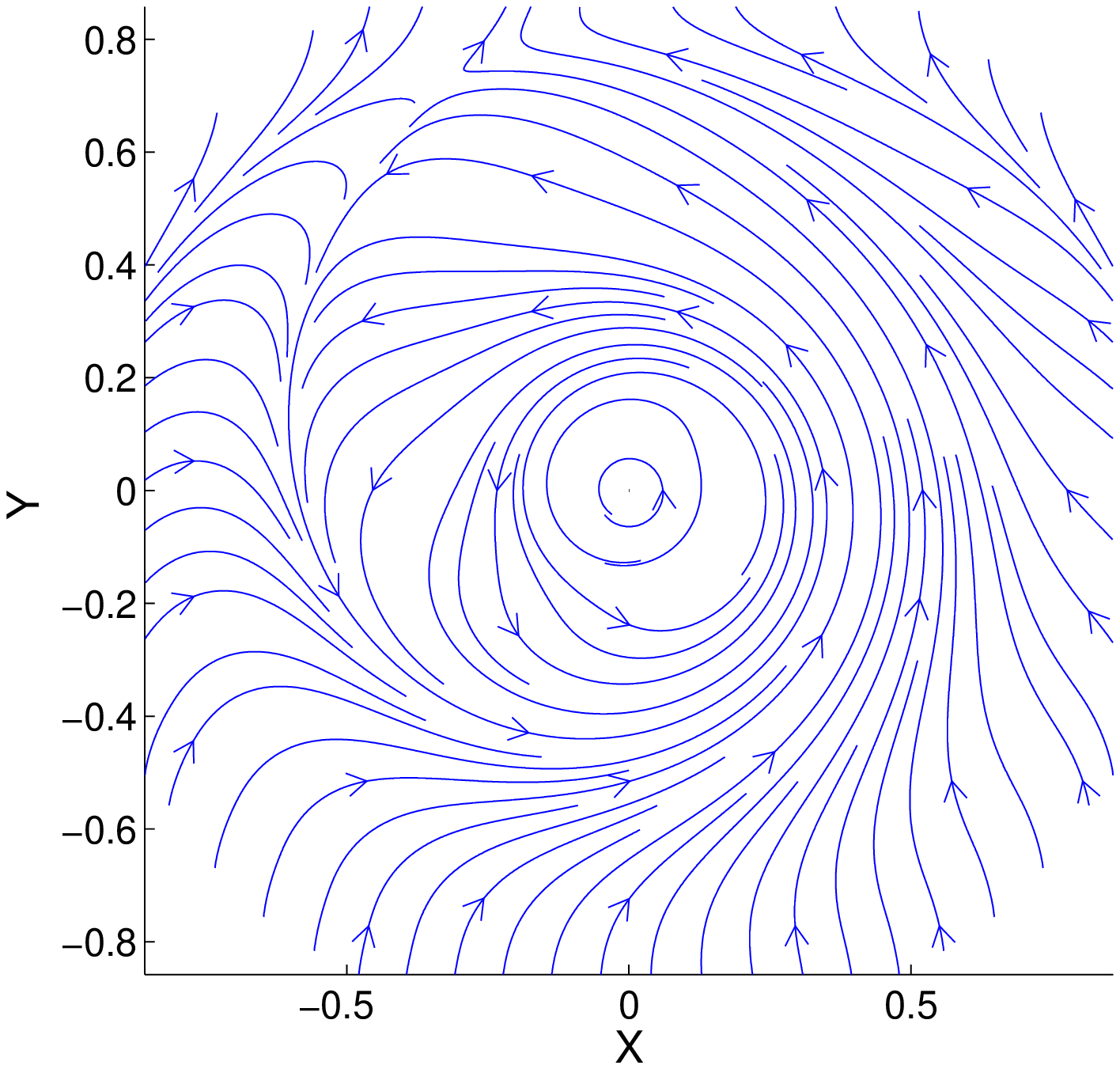}
\includegraphics[scale=0.295, trim= 8mm 0mm 0mm 0mm,clip]{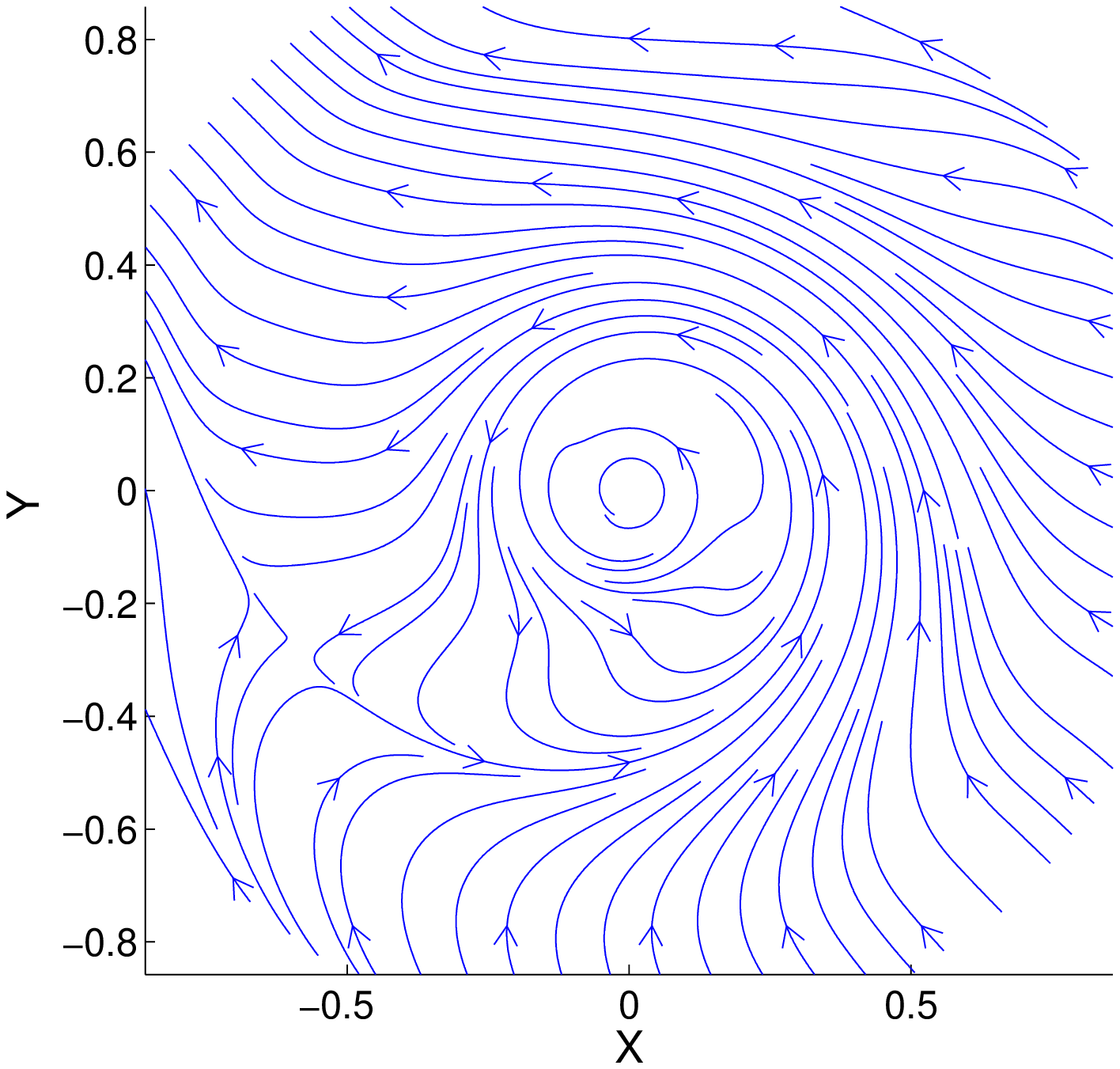}
\caption{Equatorial behaviour of the electric field projection. Case of the extreme Kerr black hole drifting 
with respect to the aligned magnetic field.}
\label{el_drift_turtle}
\end{figure}

This way we can explore also the
magnetic fields near a drifting BH, in which case the additional
electric component arises. As an example, in the right panel of
Fig.~\ref{fig2} we assumed constant velocity $\beta_y=-0.99$ directed
along the $y$-axis. A null point (again shown in the orbiting frame)
arises just above ISCO, located at $x^2+y^2=1$ for $a=1$. Finally,
Figure~\ref{fig3} shows an enlarged detail of the magnetic structure around
the region where the magnetic null point develops. The two examples
differ by the magnitude of the translatory boost of the black hole
which affects the exact location of the null point, although it always appears
very close to the horizon and requires rapid rotation of the black hole.

Finally, we remind the reader that the shape of lines of force obviously depends
on the observer's frame with respect to which the lines are plotted. Figure \ref{mag_ekv} 
compares three distinct examples. The there panels of the top row present the lines of magnetic 
force with respect to the free-falling tetrad (observer's angular momentum vanishes). 
Further down, the second row corresponds to the case of a free-falling observer 
(below the innermost stable circular orbit) in combination with a freely orbiting observer 
(Keplerian prograde rotation) above it; likewise in the third row (for retrograde rotation). 
The Schwarzschild limit $a=0$ is shown in the first column, middle column 
corresponds to spin $a=0.9$, and the last one represents the extreme case, $a=1$. 
In the bottom row the rescaled dimensionless radial coordinate 
$R\equiv\frac{r-r_{+}}{r}$ is used in order to stretch the region close to the horizon
(case of extreme rotation).

Complementary to magnetic field lines, Figure \ref{el_drift_turtle} shows the projection of
electric lines of force induced by the presence of the magnetic component by
rotation and the linear motion of the black hole. The first (left) column presents 
the non-drifting case (zero boost velocity $v_i$); in the second we set $v_x=0.5$, 
$v_y=0$; and in the third $v_x=0.5$, $v_y=0.5$. Four distinct cases are compared 
in the rows (top to bottom): zero-angular momentum observers, free-falling observers, 
co-rotating and free falling, and counter-rotating and free-falling observers, respectively. 
We observe that in all these cases the neutral electric points develop as the drift is introduced. 
We stress that as the original field is aligned ($B_x=0$) we measure the latitudinal component 
$E^{(\theta)}=0$ in the equatorial plane 
provided that $v_z=0$, so actually these plots present true shape of field lines.

We notice that non-vanishing electric component passes through the magnetic 
null, thereby accelerating electrically charged particles in this region. 
One expects reconnection to occur intermittently, as the plasma is injected 
into the dissipation region where the differently directed field lines approach
each other due to their interaction with a highly curved spacetime. Will
the gravito-magnetic effect produce the same layered structure of the
magnetic field also in the presence of non-negligible amount of plasma,
or will the field structure change  entirely? Numerical simulations will
be necessary to see whether this mechanism can be part of a broader
picture in astrophysically realistic situations, which vary wildly under
different circumstances, and to determine the actual speed at which the
process operates. We remark that, on the
other end of analytical approximations, the solution for
non-axisymmetric accretion of stiff adiabatic gas onto a rotating black hole 
also exhibits critical points near the horizon
\citep{pst88}, so the conditions for magnetic reconnection of the
frozen-in magnetic field will be again fulfilled.

\section{Conclusions}
We considered the influence of the black hole rotation acting onto the 
ordered magnetic field in the  physical frame of a star orbiting a black
hole, or plunging down  to it. If rotation is fast enough, the magnetic
layers and the corresponding null points exist just above the innermost
stable circular orbit. Although we prescribed a special configuration of the 
electro-vacuum magnetic field and we
considered only the test-field approximation, the process of warping the
field lines is a general feature that should operate also in more
complicated settings: the frame-dragging is expected to take over and
determine the field structure near the horizon. The layered structure of
the magnetic field lines with neutral points suggest this should become
a site of particle acceleration.

The essential ingredients of the scheme described here are the
{\em{}rotating black hole} and the {\em{}oblique magnetic field} into
which the black hole is embedded. The interaction region is very near
the horizon, representing, to our knowledge, the acceleration site
nearest to the black hole horizon among the variety of mechanisms
proposed so far. Even if we treated the problem in a very simplified scheme,
the idea of geometrical effects of frame dragging causing the acceleration
of matter is very promising in the context of rapidly rotating black hole
inside nuclei of galaxis.

We concentrated on the equatorial plane in which the transverse magnetic
field lines reside, but this constrain was imposed only to keep graphs
as clean as possible. Otherwise, the lack of symmetry complicates
the situation. Plasma motions in the close vicinity of the
Sgr~A* black hole are currently inaccessible to direct observation.
However, there are chances that the region will be resolved with future 
interferometers, such as GRAVITY in the near-infrared spectral band and 
the Event Horizon Telescope VLBI project in submillimeter wavelengths.
The resulting signatures in the light-curve are expected to occur synchronous 
to the X-ray signal, with time-delays specific to the radiation mechanism.

\subsection*{Acknowledgments}

We acknowledge the Czech Science Foundation (GA\v{C}R 13-00070J) and German
Forschungsgemeinschaft (DFG) collaboration project for support. We thank
an anonymous referee for helpful suggestions.

\end{document}